\documentclass[aps,prl,amsmath,amssymb,twocolumn,superscriptaddress,longbibliography]{revtex4-1}
\usepackage{graphicx}
\usepackage{braket}
\usepackage{color}
\usepackage{placeins}
\usepackage{afterpage}
\usepackage{soul}
\usepackage[normalem]{ulem}

\begin{document}


\title{Autonomous estimation of high-dimensional Coulomb diamonds from sparse measurements}

\author{Anasua~Chatterjee\textsuperscript{\dag}}
\affiliation{Center for Quantum Devices, Niels Bohr Institute, University of Copenhagen, 2100 Copenhagen, Denmark}

\author{Fabio~Ansaloni\textsuperscript{\dag}}
\affiliation{Center for Quantum Devices, Niels Bohr Institute, University of Copenhagen, 2100 Copenhagen, Denmark}

\author{Torbj$\o$rn~Rasmussen}
\affiliation{Center for Quantum Devices, Niels Bohr Institute, University of Copenhagen, 2100 Copenhagen, Denmark}

\author{Bertram~Brovang}
\affiliation{Center for Quantum Devices, Niels Bohr Institute, University of Copenhagen, 2100 Copenhagen, Denmark}

\author{Federico~Fedele}
\affiliation{Center for Quantum Devices, Niels Bohr Institute, University of Copenhagen, 2100 Copenhagen, Denmark}

\author{Heorhii~Bohuslavskyi}
\affiliation{Center for Quantum Devices, Niels Bohr Institute, University of Copenhagen, 2100 Copenhagen, Denmark}

\author{Oswin~Krause}
\affiliation{Department of Computer Science, University of Copenhagen, 2100 Copenhagen, Denmark}

\author{Ferdinand~Kuemmeth\textsuperscript{*}}
\affiliation{Center for Quantum Devices, Niels Bohr Institute, University of Copenhagen, 2100 Copenhagen, Denmark}
\affiliation{QDevil, Fruebjergvej 3, 2100 Copenhagen, Denmark}
\affiliation{\textsuperscript{*}kuemmeth@nbi.dk\\\textsuperscript{\dag}These authors contributed equally to this work.}

\newcommand{\VL}{V_\mathrm{L}}
\newcommand{\VM}{V_\mathrm{M}}
\newcommand{\VR}{V_\mathrm{R}}

\newcommand{\Btot}{B^\mathrm{tot}}
\newcommand{\Bext}{B^\mathrm{ext}}
\newcommand{\Bznuc}{B_\mathrm{z}^\mathrm{nuc}}
\newcommand{\Bpnuc}{B_\perp^\mathrm{nuc}}

\newcommand{\ud}{\uparrow\downarrow}
\newcommand{\du}{\downarrow\uparrow}

\newcommand{\drv}{\mathrm{d}}

\newcommand{\FHahn}{F_\mathrm{Hahn}}
\newcommand{\FCPMG}{F_\mathrm{CPMG}}
\newcommand{\FFID}{F_\mathrm{FID}}
\newcommand{\Fenv}{F_\mathrm{env}}

\newcommand{\Ga}{^{69}\mathrm{Ga}}
\newcommand{\Gb}{^{71}\mathrm{Ga}}
\newcommand{\As}{^{75}\mathrm{As}}
\newcommand{\fGa}{f_{^{69}{\rm Ga}}}
\newcommand{\fGb}{f_{^{71}{\rm Ga}}}
\newcommand{\fAs}{f_{^{75}{\rm As}}}

\newcommand{\TCPMG}{T_2 ^\mathrm{CPMG}}
\renewcommand{\vec}[1]{{\bf #1}}

\newcommand{\remove}[1]{{\color{red} \sout{#1}}}
\newcommand{\add}[1]{{\color{blue} #1}}
\newcommand{\comFK}[1]{{\color{cyan} #1}}
\newcommand{\com}[1]{{\color{green} #1}}

\date{\today}

\begin{abstract}
Quantum dot arrays possess ground states governed by Coulomb energies, utilized prominently by singly occupied quantum dots, each implementing a spin qubit. For such quantum processors, the controlled transitions between ground states are of operational significance, as these allow the control of quantum information within the array such as qubit initialization and entangling gates. For few-dot arrays, ground states are traditionally mapped out by performing dense raster-scan measurements in control voltage space. These become impractical for larger arrays due to the large number of measurements needed to sample the high-dimensional gate-voltage hypercube and the comparatively little information extracted. We develop a hardware-triggered detection method based on reflectometry, to acquire measurements directly corresponding to transitions between ground states. These measurements are distributed sparsely within the high-dimensional voltage space by executing line searches proposed by a learning algorithm. Our autonomous software-hardware algorithm accurately estimates the polytope of Coulomb blockade boundaries, experimentally demonstrated in a 2$\times$2 array of silicon quantum dots.
\end{abstract}

\maketitle

\begin{figure*}
\includegraphics[scale=0.2]{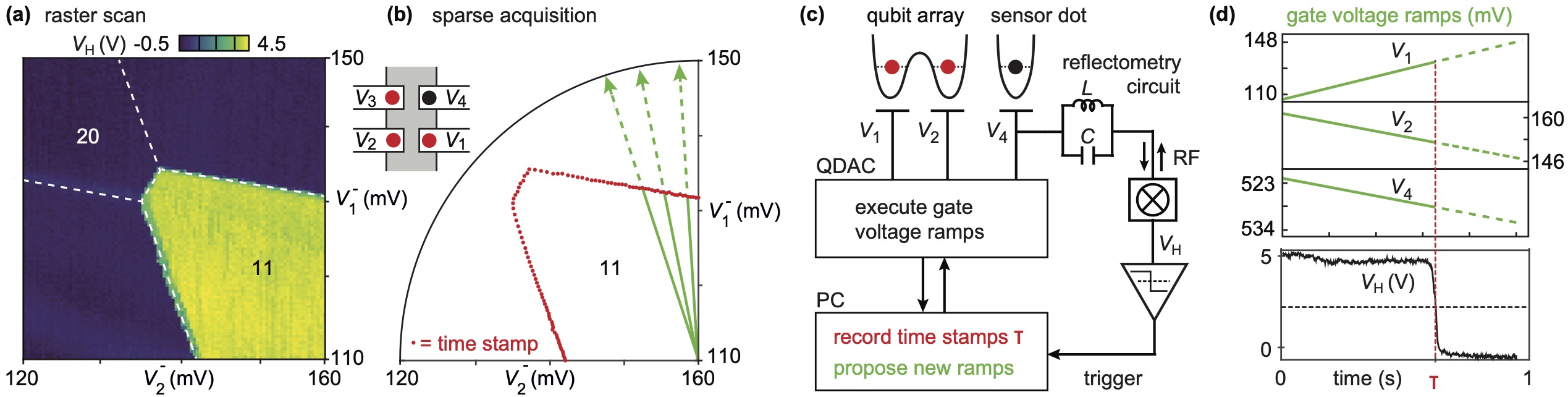}
\caption{Principle of sparse acquisition, replacing here a 100$\times$100 raster of measurements by the acquisition of 100 time stamps, which we extend to higher dimensions. 
(a) Typical raster scan of a double-dot-with-sensor stability diagram, requiring point-by-point gate-voltage steps and associated pixel-by-pixel acquisition of the sensor dot signal $V_\mathrm{H}$. 
Inset: schematic of the 2$\times$2 quantum dot device, allowing up to three qubit dots (red) and one sensor dot (black) controlled by $V_\mathrm{1,2,3,4}$.
Negative superscripts of control voltages $V_\mathrm{1,2,3}$ indicate that $V_\mathrm{4}$ compensates the sensor dot against cross talk from gates 1, 2, and 3 (see main text).
(b) Sparse acquisition of the same stability diagram using time stamps. 
Line searches are continuously executed in gate-voltage space (green rays) radially outward from inside 11, creating time stamps (red dots) whenever leaving the 11 ground state. Here, 11 indicates one-electron occupation for dot 1 and dot 2. 
(c) Searches in high-dimensional voltage space are implemented as synchronized ramps on all QDAC channels, shown here for $V_1^-$ and $V_2^-$. 
Whenever the charge configuration of the qubit dots changes, $V_\mathrm{H}$ changes abruptly, triggering the acquisition of a time stamp.
(d) QDAC voltages (green) and $V_\mathrm{H}$ (black) for one exemplary search in panel b. Except for this illustration, $V_\mathrm{H}(t)$ is not recorded.   
After a trigger event, the QDAC aborts the remainder of the search (green dashed segment) and proceeds with the next. 
}
\label{fig1}
\end{figure*}

\section{Introduction}
Advances in the engineering of high-quality spin qubits \cite{Yoneda2018,Muhonen2014,Watson2018,Yang2020} have brought the field closer to new bottlenecks along the path to scaling to larger capacitively-coupled arrays \cite{Zajac2016, Volk2019, Mills2019, Lawrie2019, Dehollain2020,Ansaloni2020,Mortemousque2020,vanRiggelen2021,Fedele2021}. 
Among these is the challenge of navigating the high-dimensional control-voltage space that is associated with gate-controlled spin qubits~\cite{Dehollain2020,Volk2019,Ansaloni2020,Mills2019}. 
In such devices, many gate voltages have to be carefully tuned to find transitions between specific ground states that control loading and unloading of electrons in the array as well as wavefunction overlap between dots during coherent qubit operations.

Recent attempts to automatically tune quantum dot systems \cite{Zwolak2020,Zwolak2021,Moon2020,Lennon2019,Kalantre2019,Botzem2018,Darulova2020,Ziegler2022} focused on implementing algorithms on the software level, while mostly maintaining the paradigm of two-dimensional raster scans as input. 
While this has historically been a successful technique for small devices, the increasing number of physical gate electrodes in quantum dot arrays increases the dimensionality of control parameter space, where dense sampling is neither conducive (data cannot be visualized easily for human assessment) nor time efficient. For example, using traditional measurements the sequential acquisition of 100$^8$ pixels associated with a hypercube of 8 gate electrodes would require many years.
These scaling challenges may be alleviated by automated and smart navigation algorithms that make use of high-bandwidth measurement techniques to quickly navigate within high-dimensional parameter spaces without the need for human intervention.  

A fundamental prerequisite for spin-qubit operations in weakly tunnel-coupled quantum dots is the existence of large volumes in parameter space (so-called Coulomb diamonds) in which the ground-state occupation of the quantum-dot array is well-defined and stable due to Coulomb blockade, as well as the identification of boundaries between one ground-state configuration and another. For example, the boundary between 01 and 11 of a double dot is useful for resetting the left spin, whereas the boundary between 11 and 20 is useful for inducing two-qubit rotations (Heisenberg spin-exchange rotations), spin teleportation, or spin-to-charge conversion (qubit readout)~\cite{Chatterjee2021}. Here, each digit shows the occupancy of a quantum dot.
For larger quantum-dot arrays, measuring an entire high-dimensional gate-voltage hypercube by means of raster scans just to locate operational facets would not be feasible, as it results in high costs in terms of data storage and time spent in measuring uninformative regions of gate space. 
In machine-learning applications especially, the presence of measurement cost often leads to the deploy of active-learning algorithms to acquire only those samples that are most informative for training the model; it is therefore important to discard irrelevant data early on in the algorithmic procedure. Additionally, the crosstalk between gate electrodes and quantum dots may become an operational burden for larger arrays~\cite{Heinz2021}. One approach to mitigate cross talk in small arrays has been the use of linear combinations of gate voltages (sometimes called virtual gates)~\cite{Volk2019,Mills2019}, yet measurements are typically still performed as pairwise 2D maps of such control parameters to avoid dense sampling.

In this work, we present a two-pronged automation that comprises (A) device-led hardware triggering and (B) active learning to autonomously estimate Coulomb facets (ground-state boundaries) from a small number of triggered measurements. We start with (A) and show that reflectometry, tuned to be sensitive to charge state boundaries in a binary manner, allows sparse measurements triggered by the device itself, rather than the control software, removing the constraint of raster scans and opening up to new algorithmic acquisition procedures. By monitoring a high-bandwidth sensor dot via high-frequency reflectometry in a 2$\times$2 array of silicon quantum dots \cite{Ansaloni2020,Bohuslavskyi2020}, we map a device property of interest (a particular ground state, for example) onto the sensor signal. Gate voltages are ramped in user-defined directions, however, a data point is only acquired when a charge-state boundary is encountered. When this happens, the sensor signal sharply changes, triggering the acquisition of a time stamp that later is decoded to gate voltages. We illustrate this technique by configuring the device as a double-dot with sensor dot, although it is not limited to such a device. We then turn towards (B) and show that such sparse acquisitions can directly be analyzed by an active-learning algorithm, which adaptively proposes new ramp directions. We demonstrate that our algorithm can autonomously estimate Coulomb blockade boundaries based on a small number of algorithm-based sparse acquisitions, revealing the Coulomb facet of a quadruple dot in four-dimensional gate-voltage space.

\section{Device-triggered time stamps}
A typical raster scan of a 2D quantum dot stability diagram requires setting the requisite gate voltages step-by-step, and a query and acquisition of a signal for each pixel (Fig.~\ref{fig1}a). 
The signal is typically a current through a double dot or, in our case, the demodulated reflectometry signal $V_\mathrm{H}$ arising from a proximal sensor dot~\cite{Volk2019a}. Sensitivity of $V_\mathrm{H}$ to state boundaries (charge rearrangements within the 2$\times$2 array) is achieved by operating the sensor dot at degeneracy with its reservoir (tunneling between dot 4 and the reservoir then results in a maximal value of $V_\mathrm{H}$), and by negatively compensating the sensor dot 4 gate $V_\mathrm{4}$ such that changes applied to the other three dot gates $V_\mathrm{1,2,3}$ do not affect the potential of dot 4 (the device and measurement setup is described in previous work~\cite{Bohuslavskyi2020} as well as in Appendix A).

Rastering results in the acquisition of a great many samples of no added value (bulk of the blue and yellow regions in~\ref{fig1}a). In addition, a significant proportion of ``dead time'' is needed for communication with instruments controlling gate voltages and for reading sensor voltages at each pixel of the raster scan. For spin-qubit applications, it is the low-dimensional boundary (white dashed line in panel a) that is of interest (for example for inducing exchange oscillations at the 11-20 boundary) as elsewhere spin-qubit dynamics is effectively suppressed by Coulomb charging energies. 

In our triggered acquisition, gate voltages are not stepped pixel-by-pixel but continuously ramped in an appropriate pattern (Fig.~\ref{fig1}b uses radial polar scans) or as dictated adaptively by a learning algorithm (see below). 
No data is recorded during ramping. When a charge state boundary is encountered (Fig.~\ref{fig1}c), the sensor signal $V_\mathrm{H}$ acts as a digital trigger that prompts the recording of a time stamp $\tau$ (Fig.~\ref{fig1}d), after which the ramp generator (QDevil QDAC~\cite{QDevil}) aborts the remaining segment of the search (green dashed line) and starts executing the next search. 
Once all searches are executed, the gate voltages associated with each time stamp are plotted (red dots in panel b).

Below, we apply device-triggered time stamping to a quadruple dot in which one dot serves as a sensor dot (Fig.~\ref{fig3}), and combine it with an active-learning algorithm that estimates the same Coulomb facet from even fewer time stamps (Fig.~\ref{fig4}).

\begin{figure}
\includegraphics[scale=0.2]{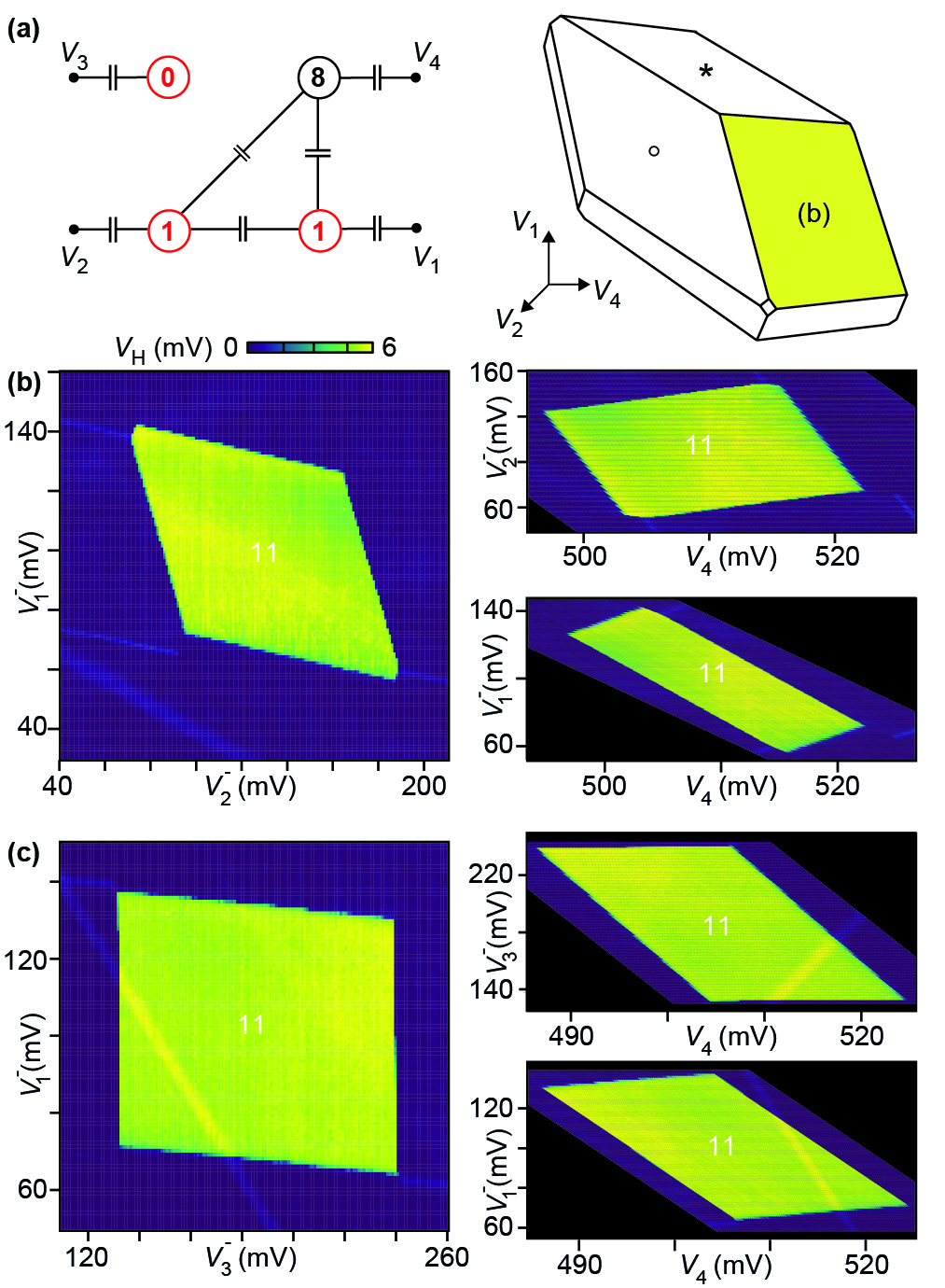}
\caption{Effect of capacitances and projection on facet shapes.  
In Figure~\ref{fig1}a, gate voltages that correspond to charge degeneracy of the sensor dot (yellow pixels) can be modeled as the facet of a triple-dot Coulomb diamond, i.e. as a 2D manifold in 3D voltage space $V_\mathrm{1,2,4}$. 
(a) Simulated Coulomb diamond of a triple dot with small capacitance between dot 2 and dot 4. The facet corresponding to the addition of an electron to dot 1 is tetragonal ($\star$), whereas facets adding an electron to dot 2 ($\circ$) or dot 4 (yellow) are hexagonal. (b) Measurements of the dot-4 facet projected onto any pair of $V_\mathrm{1,2,4}$, revealing three distorted hexagons. (c) Same as (b), but activating dot 3 instead of dot 2 to realize a triple dot with a tetragonal dot-4 facet. 
Generalizing these observations to a quadruple dot, we expect that facets are 3D manifolds in 4D voltage space, with different facets having different shapes. One such facet, projected to 3D, is the topic of Figure~\ref{fig4}c.}
\label{fig2}
\end{figure}

\subsection{Triple-Dot Facet}

The device configuration associated with Fig.~\ref{fig1}a can be modeled by a triple-dot capacitance circuit shown in Fig.~\ref{fig2}a. 
Simulations of such circuits based on the constant interaction model~\cite{vanderWiel2002} reveal ground-state regions that extend in $V_\mathrm{1,2,4}$ space. For a triple-dot circuit with three-fold symmetry in the capacitance values, one expects a three-fold symmetric region bounded by 12 sides. For the circuit considered here, the smaller capacitance between dot 2 and dot 4 breaks that symmetry, and we obtain 14 sides as shown. From this simulation, we expect that the facet adding an electron to dot 1 is tetragonal, while facets adding an electron to dot 2 or dot 4 are hexagonal.

Experimentally, this is confirmed in Fig.~\ref{fig2}b by a conventional raster scan of $V_1$ vs. $V_2$, with $V_4$ compensated in a linear fashion. 
We observe a \emph{hexagonal} dot-4 manifold and plot its two-dimensional projections onto $V_\mathrm{1}$-$V_\mathrm{2}$, $V_\mathrm{2}$-$V_\mathrm{4}$, and $V_\mathrm{1}$-$V_\mathrm{4}$. 
In the yellow region marked as 11, dot 1 and dot 2 are singly occupied. 

Analogous confirmation of the \emph{tetragonal} dot-1 facet would require reflectometry off gate 1, which we did not implement. Instead, activating dot 3 instead of dot 2 realizes in the same device a geometrically different triple dot with a different set of capacitances, while keeping dot 4 as a sensor dot (Fig.~\ref{fig2}c). Here, the dot-4 facet is tetragonal, as expected from simulations~\cite{Bohuslavskyi2020}.

Note that plotting lower-dimensional projections of facets --- here projecting onto two dimensions 2D manifolds that are embedded in 3D voltage space --- simplifies visualization but does not change the physics associated with different facet shapes. Accordingly, the different projections in Fig.~\ref{fig2}b (and similarly in~\ref{fig2}c) show equivalent versions of the same facet, with distortions that quantitatively depend on the capacitive crosstalk  between gate electrodes and nearby dots.

\begin{figure*}
\includegraphics[scale=0.18]{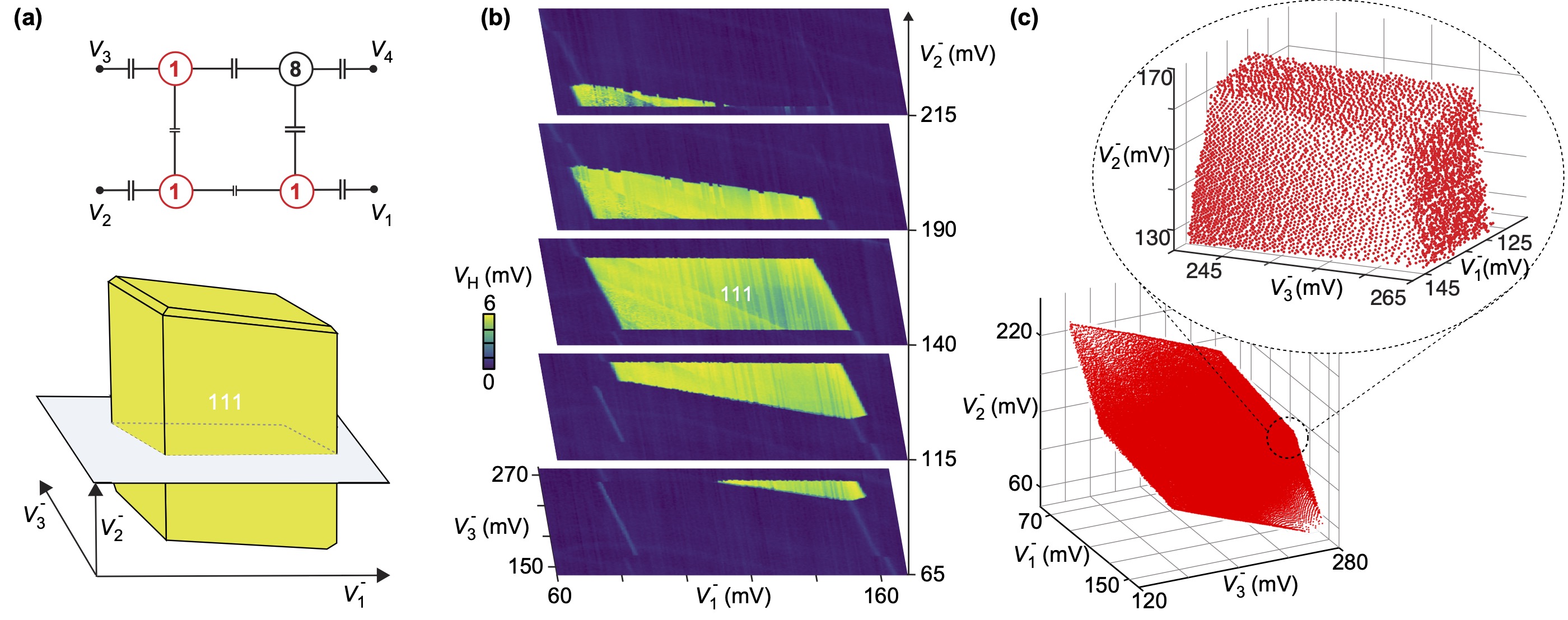}
\caption{Facet of a quadruple-dot Coulomb diamond. 
(a) Capacitance model of a quadruple dot. The facet corresponding to the addition of an electron to dot 4 is a 3D manifold in 4D voltage space, projected here onto $V_\mathrm{1,2,3}^-$ (yellow volume). 
(b) Measured 2D cuts through the dot-4 facet at fixed values of $V_\mathrm{2}^-$. 
(c) Same facet as in (b), but acquired by orienting 100,000 line searches uniformly in $V_\mathrm{1,2,3}^-$ space. 
Inset: density of time steps near a significant corner of this manifold (see main text). }
\label{fig3}
\end{figure*}

\subsection{Quadruple-Dot Facet}

Spin qubits have been encoded in single quantum dots (Loss-DiVincenzo encoding), double dots (singlet-triplet encoding), and triple dots (exchange-only encoding)~\cite{Chatterjee2021}. 
To accomodate a sensor dot, the typical implementation of an exchange-only qubit is a quadruple dot~\cite{Medford2013, Eng2015, Malinowski2017b}. 
To explore their high-dimensional facets, raster scans come up short. 
While charge stability diagrams of double dots are well established~\cite{vanderWiel2002} and triple dots have been studied by brute-force raster scans~\cite{Granger2010}, the charge stability diagrams of quadruple dots are largely unexplored~\cite{Bohuslavskyi2020}.

Analogously to the triple dot, for the fully energized 2$\times$2 array, simulations show that facets are 3D manifolds embedded in 4D gate-voltage space, with shapes that depend on details of the underlying quadruple-dot capacitance circuit. For the purpose of visualization, we also expect that it suffices to show one projection of a facet from four-dimensional space onto three dimensions. For the simulation in Fig.~\ref{fig3}a, we choose the projection onto $V_\mathrm{1}$, $V_\mathrm{2}$ and $V_\mathrm{3}$ and observe a dot-4 facet that is bounded by 10 sides (yellow region labeled 111, in which dot 1, dot 2 and dot 3 are all singly occupied). While this illustrative simulation is not intended to be quantitative, it deliberately breaks the four-fold symmetry of the capacitance circuit by assuming interdot capacitances that are larger for dot 4 (which in the experiment is occupied by 8-9 electrons) compared to the other, singly-occupied dots. 
For identical capacitances, the dot-4 facet would be invariant under exchange of $V_\mathrm{1}$, $V_\mathrm{2}$ and $V_\mathrm{3}$, and the simulated projection then shows indeed a three-fold symmetric facet bounded by 12 sides.

Figure~\ref{fig3}b shows several conventional raster scans $V_\mathrm{1}^-$ vs. $V_\mathrm{3}^-$ for five values of $V_\mathrm{2}^-$. 
For these measurements, the quantum dot array is tuned as a quadruple dot by activating all four gate electrodes. 
The charge degeneracy between 1118 and 1119 is clearly revealed (yellow region labeled 111), demonstrating the sensitivity of the resulting cross section to $V_\mathrm{2}^-$ and hence the inadequateness of traditional characterization by a 2D raster. 

On the other hand, Figure~\ref{fig3}c shows the same manifold as explored in~\ref{fig3}b, but acquired sparsely by time stamping. 
Specifically, a Fibonacci grid, commonly used to homogeneously sample the surface of a sphere, is used to distribute 100,000 hardware line searches in three-dimensional polar coordinates associated with $V_\mathrm{1,2,3}^{-}$. After converting all 100,000 time stamps to gate voltages $V_1$, $V_2$, $V_3$, $V_4$, we omit $V_4$ to plot a three-dimensional projection, as in panel ~\ref{fig3}a.
The number of pixels shown in~\ref{fig3}b is (210$\times270)\times5\approx$ 0.3M, with each 2D raster taking approximately 10 minutes including computer-instrument communication and data transfer. We note that a naive 300$^3$ rastering of the $V_\mathrm{1}^-$,$V_\mathrm{2}^-$,$V_\mathrm{3}^-$ hypercube would require several days, during which slow sensor drifts may further complicate measurements unless devices are perfectly stable. The line-search method used in~\ref{fig3}c provides a more detailed and efficient characterization compared to brute-force rastering, taking roughly six hours. A 300$^4$ rastering of the entire $V_\mathrm{1}$,$V_\mathrm{2}$,$V_\mathrm{3}$,$V_\mathrm{4}$ hypercube would require several years.\\

\section{Active-learning-assisted sparse measurements}

\subsection{Active-Learning Algorithm}
The choice of executing 100,000 predetermined and uniformly distributed searches for Fig.~\ref{fig3}c was motivated by our desire to not miss potentially small features of this unknown facet while acquiring significantly fewer samples than a $100^3$ pixel cube. Motivated by the simple geometric shapes associated with an arbitrary number and geometry of capacitively coupled quantum dots, namely convex polytopes for ground states of the constant interaction model~\cite{nazarov2009quantum, Krause2021}, we now turn towards an even more drastic reduction of acquisitions by estimating the dot-4 facet from a small number of line searches, adaptively oriented using machine-learning techniques.

In brief, the key idea is that as the algorithm learns the locations of more and more time stamps in gate space, it can estimate the unknown polytope and suggest new search directions to efficiently reject, improve, or validate its best estimate. Once a polytope is validated by the acquisition of new time stamps, the algorithm terminates and returns the polytope. Even though we demonstrate this algorithm on the 2$\times$2 device, we note that the algorithm and its implementation is not limited to quadruple dots. The algorithm is described further in Appendix B and in technical detail in~\cite{Krause2021}. 

For illustrative purposes, however, we first explain the algorithm in Fig.~\ref{fig4}a using a two-dimensional problem (appropriate for a device configuration as in Fig.~\ref{fig4}b, in which dot 3 is kept in Coulomb blockade), before we apply it to estimate the dot-4 facet of the full 2$\times$2 device (Fig.~\ref{fig4}c).  

Figure~\ref{fig4}a illustrates the iterative nature of the algorithm. To accommodate experimental noise, i.e. uncertainty in the exact gate voltages associated with a real charge transition, the algorithm associates with each hardware-triggered time stamp a confidence interval. 
Its end points (blue and orange points in panel~\ref{fig4}a) are assumed to straddle the unknown boundary.  
This not only allows mitigation of experimental noise in the triggering apparatus or non-ideal device characteristics, but, importantly, replaces the polytope estimation problem by a binary large-margin classification problem, namely to find planes that optimally separate the inner points from the outer points. 
By combining a second-order solver for the large-margin problem with an active-learning scheme for repeated verification~\cite{Krause2021}, the algorithm iteratively and efficiently arrives at an accurate estimate of the actual charge state boundary. 

For the results presented here, a nominal ramp speed of 3 s per line search was used. In practice, each search takes significantly less time since it is terminated and followed by the next search as soon as a trigger event occurs. Equivalent results were obtained by reducing the nominal ramp time from 3 s to 0.5 s, and further reductions are realistic due to the high bandwidth associated with $V_\mathrm{H}(t)$. 
We observed a signal-to-noise ratio of 2.3 for a $V_\mathrm{H}(t)$ rise time of 10 $\mu$s~\cite{Bohuslavskyi2020}, but did not yet take advantage of this speed due to the heavy low-pass filtering in the cryostat control wiring for $V_\mathrm{1,2,3,4}$. 

After several such iterations, once all sides of the estimate lie within confidence intervals of the new points, the estimate is considered to match the ground truth (yellow region in Fig.~\ref{fig4}a) and returned to the user.

\subsection{Active-Learning Results}
\begin{figure}
\includegraphics[scale=0.2]{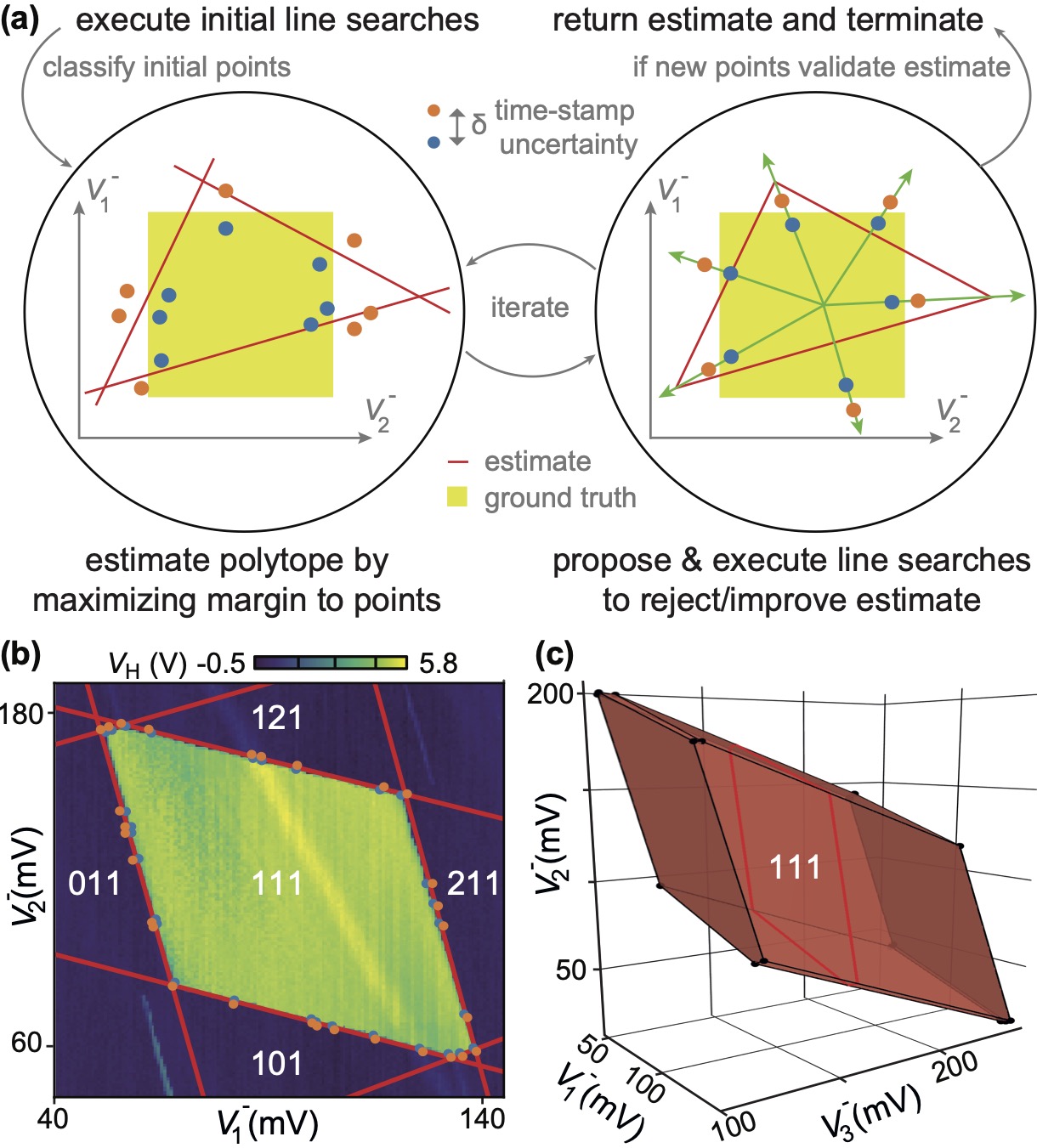}
\caption{
Autonomous estimation of high-dimensional facet polytopes from sparse time stamps.  
(a) Active learning in high-dimensional voltage space, illustrated here for $V_1^-$ and $V_2^-$. Time stamps are generated by algorithm-controlled line searches and converted to confidence intervals (blue-red points) to allow classification via a large-margin classifier. 
Based on randomized initial searches, the algorithm makes a best estimate of the polytope (red triangle). Subsequent line searches are then directed towards the corners of the estimate and towards the middle of each boundary, to reject or validate it (green arrows). Estimation and new line searches iterate until consecutive searches result in the same estimate, within tolerance, which then represents an accurate model of the ground truth (yellow square).
(b) Algorithm applied to $V_1^-$ and $V_2^-$, terminating with an estimate (six red lines) based on 29 time stamps (blue-orange pairs). For comparison, a conventional 150$\times$150 pixel raster scan is also shown ($V_\mathrm{H}$). 
(c) Dot-4 facet estimated by the algorithm for the qubit array in the 111 state, plotted here as a projection onto $V_\mathrm{1,2,3}$ (10 red planes).  
For clarity, the intersections of planes are highlighted (black lines and dots) as well as a hexagonal cross section for constant $V_\mathrm{3}$ (red line). }
\label{fig4} 
\end{figure}

To demonstrate the drastic reduction in the number of required acquisitions, we first wish to learn a facet similar to the one in the 22100-pixel raster scan of Fig.~\ref{fig2}b. We do this by fixing $V_3$ in Coulomb blockade of its first electron, i.e. the dot-4 facet then corresponds to the degeneracy between 1118 and 1119, and by controlling $V_1^-$ and $V_2^-$ via the algorithm. 
After execution of only 56 line searches, the algorithm terminates after a minute with the estimate shown in Fig.~\ref{fig4}b (red lines), along with the associated set of confidence intervals (29 blue-orange pairs). 
We also overlap a conventional raster scan measured independently in the same region, indicating that the algorithm accurately determined the correct polytope from a sparse set of time stamps. 

Similar tests that also included simulated charge stability diagrams and randomly generated polytopes did not reveal any signs of bad local optima~\cite{Krause2021}. 
Likely, this is a consequence of our iterative active-learning scheme, which by design not only verifies the estimated polytope, but also searches for potential modeling errors using new line searches. Thus, a local optimum likely gets caught and subsequently mitigated by a new training iteration. 
This may explain the high density of time stamps near small features of Figure~\ref{fig4}b. Notabably, the algorithm managed to find two small sides, corresponding to qubit-dot transitions from 111 to 021 and 201.
These interdot transitions are barely visible on the raster scan, but are of significant operational value for example for the coherent control of exchange-only qubits~\cite{Gaudreau2011, Medford2013, Eng2015, Malinowski2017b}. 

Finally, we let the algorithm control $V_1^-$, $V_2^-$ and $V_3^-$, in order to estimate the dot-4 facet of the fully activated quadruple dot. 
After execution of a total of 280 line searches, the algorithm terminates after 12 minutes with the estimate shown in Fig.~\ref{fig4}c (10 red planes). 

Our results show that it is possible to autonomously estimate the polytope of charge-state transitions within a quantum dot array in a real experimental setting. Overall, the algorithm returns an accurate estimate of the true polytope volumes and shapes, including small features, from a small number of hardware line searches. Comparing the performance of our algorithm to an idealized baseline algorithm that can distinguish different neighboring charge states, we find that the majority of errors is introduced by the difficulties of solving the NP-hard polytope estimation problem, rather than our active-learning scheme~\cite{Krause2021}. 
Thus, our method will likely profit from continued development of improved and faster estimation algorithms. 

Our results also motivate the creation of additional algorithmic layers that build on top of autonomous estimation of charge stability diagrams. For instance, the output of our algorithm (Coulomb polytopes) could be analyzed by a meta algorithm to extract various device parameters, such as mutual capacitances that then serve as input for the automated definition of ``virtual gates"~\cite{Hsiao2020}. Another opportunity is to increase the dynamic range of the sensor signal, by broadening the sensor response or placing the sensor dot further away~\cite{Volk2019a}, such that the algorithm not only learns the boundary of a target state, but also to distinguish different neighboring charge states (white numbers in Fig.~4b). Our baseline analysis indicates that such a non-binary sensor signal would make it easier to identify the polytope with high accuracy and at least an order of magnitude faster. This may open a path towards automated abstraction of physical gate operations, such as loading  or shuttling individual electron spins or exchange-coupling specific pairs.  Even for devices that are available today with similar readout signal-to-noise ratio and control complexity~\cite{Volk2019, Borjans2021}, we foresee an impact on their tuning and operational functionalities in the sense that human expert tuning capabilities are no longer the limiting factor~\cite{Moon2020}.

\section{Discussion and Conclusion}
In this work, we presented two synergistic advancements with potential applications to the tuning of large quantum dot arrays and spin-qubit processors, constituting the first automatic discovery of state transitions in the literature.
The first is a hardware-based sparse acquisition technique, based on high-fidelity sensing by reflectometry that enables digital detection of quantum dot charge states, as well as the ability to perform ramped line searches, freeing the measurement process from a reliance on raster scans and moving to device-led measurements. This has enabled the second advancement, an adaptive machine-learning algorithm, here an unsupervised acquisition process that finds the high-dimensional shape of charge state boundaries in capacitively-coupled quantum dot arrays. 

A future variation of our line searches may involve acquisition of corresponding reflectometry traces (similar to the $V_\mathrm{H}(t)$ trace shown in Fig.~\ref{fig1}d). With appropriate digitizers, memory, and real-life signal processing, such traces may allow determination of boundaries even if device properties are less ideal, for example if the sensor signal is too noisy for threshold triggering or if imperfect sensor compensation results in threshold drifts.  

A primary application of our sparse acquisition technique, as illustrated in Figure~\ref{fig3}c, is to gather relevant information about multi-dot configurations that may be difficult to obtain in reasonable timescales from naive raster scans. In addition, as larger and denser arrays of quantum dots are fabricated, it will become imperative to quickly initialize and operate them via knowledge of their high-dimensional ground-state boundaries. Many of the involved steps, such as calibrating the sensor's operating voltages and capacitive cross coupling, can be automated such that quantum-dot arrays can be initialized and charge-state boundaries be recognized autonomously. In addition, the high signal-to-noise ratio of our high-bandwidth sensing signal (see for instance $V_\mathrm{H}(t)$ in Figure~\ref{fig1}d) should allow orders of magnitude faster ramps compared to our demonstration, thereby making the automatic recognition of facets and automated tune-up of multi-dot arrays more feasible, perhaps even in novel non-ideal materials with slow charge fluctuations. Finally, by allowing the device to trigger a measurement in regions of most information, the risk of expert human intuition restricting itself to regions of known physics is eliminated, with the result that interesting operational regions may be discovered and utilized. 

For example, perfect corners such as in the zoom of Fig.~\ref{fig3}c may be useful for permutational quantum computing ~\cite{Jordan2010,Ansaloni2020}, whereas hard-to-find ground-state boundaries such as the smallest side visible in Fig.~\ref{fig2}a correspond to the simultaneous tunneling of multiple electrons at different locations of the array~\cite{Gaudreau2006, Hamo2016, Bohuslavskyi2020}. For quadruple dots, such correlated charge dynamics may be utilized as quantum cellular automata~\cite{Lent1993}, with possible applications as digital logical gates~\cite{Amlani1999} or quantum
registers~\cite{toth2001}.

Going towards larger and larger quantum-dot arrays~\cite{Krause2022}, the number of facets increases exponentially. For example, the number of primary facets alone, i.e. the different ways of adding or removing one electron from an array of $n$ dots, scales as $2^n$. 
In addition, ground-state transitions that occupy vanishingly small volumes in parameter space emerge, such as the correlated motion of multiple electrons changing their position simultaneously, much like in large quantum cellular automata~\cite{Lent1993}, making it more and more difficult to reliably find such transitions in the presence of experimental noise. 
Even in the absence of noise, the estimation of a ground-state polytope from samples of its boundary becomes NP-hard~\cite{Krause2021}. 
 
However, for the tune-up of a large spin-based quantum processor, one may only be interested in a small number of electron movements, such as initializing each dot with one electron or finding pairwise inter-dot transitions for two-qubit exchange gates. In practice, this might be possible by operationally partitioning the device in blocks of smaller qubit arrays, each read out individually by a dedicated sensor, to estimate the polytopes associated with the smaller arrays rather than the entire system. 

For example, the fact that our algorithm also works in 5-dimensional parameter space~\cite{Krause2021,Krause2022} allows the implementation of an infinite linear chain of quantum dots, in which all local transitions of any dot can be estimated, within the two-local neighborhood of the dot, from the signals of sensors located parallel to the chain. 
The optimum size of such qubit blocks, and the efficiency of our algorithm in this approach, needs to be investigated further, taking into account the qubit-qubit connectivities and cross talk of realistic future devices. A recent proposal aims at avoiding dense qubit arrays altogether, by bringing pairs of electrons near each other only when required~\cite{Boter2021}, but may still require efficient learning of charge stability diagrams. 

Our techniques, therefore, show promise for operating interconnected quantum dot arrays, where the parameter space extends beyond three dimensions and  established small systems, i.e. beyond the visualization and control experience of human intuition. This includes intermediate-size qubit arrays with multiple nearest-neighbor connectivities for quantum simulation, and, ultimately, scalable lattices suitable for quantum error correction \cite{Terhal2015,Boter2021} and universal quantum computing. 

\section{Acknowledgements}
We thank Silvano De Franceschi and Louis Hutin for providing the silicon device. This work received funding from EU grant agreements No. 688539, 856526 and 951852. AC acknowledges support from the Independent Research Fund Denmark. A.C. and F.A. contributed equally to this work.

\appendix

\section{Appendix A: Device and Measurement Setup}

To demonstrate our technique in a real setting, we apply it to a foundry fabricated transistor-like device at base temperature of a dilution refrigerator. 
The channel consists of an undoped silicon nanowire (70~nm wide and thickness 7~nm thick) connected to highly doped source and drain reservoirs. A 2$\times$2 array of quantum dots is formed under four accumulation gates (each 32~nm long and spaced by 32~nm, see Fig.~1b inset and Refs.~\cite{Ansaloni2020,Bohuslavskyi2020} for details). 
Throughout the manuscript, we refer to dot 4 as the sensor dot and consider the other, singly-occupied dots as qubit dots. 

To perform reflectometry, a 191~MHz carrier is applied to a resonant tank circuit that is wirebonded to gate 4, and its reflected wave is converted to a voltage $V_\mathrm{H}$ by homodyne demodulation~\cite{Volk2019a}. 
This signal is amplified (Stanford Research SR560) such that it can serve as a TTL-level trigger for the recording of a time stamp, after which the next ramp is started. A digital multimeter (Keysight 34465A) is used to detect the trigger and save as a time stamp. The systematic delay between voltage changes generated by the QDAC and voltage changes arriving on the gate electrodes (due to RC filtering in the setup) is small (\textless10 $\mu$s) and is not taken into account.
Sensitivity of $V_\mathrm{H}$ to state boundaries (charge rearrangements within the 2$\times$2 array) is achieved by operating the sensor dot at degeneracy with its reservoir (tunneling between dot 4 and the reservoir then results in a maximal value of $V_\mathrm{H}$), and by negatively compensating $V_\mathrm{4}$ such that changes applied to $V_\mathrm{1,2,3}$ do not affect the potential of dot 4~\cite{Bohuslavskyi2020}. 
We indicate the presence of this linear $V_\mathrm{4}$ compensation by a negative superscript for the control voltages, $V_\mathrm{1,2,3}^-$.  
Within a charge state, $V_\mathrm{H}$ is then independent of the other gate voltages, but changes abruptly whenever an electron movement occurs in the array, i.e whenever a ground-state to ground-state transition is encountered. 

This process is illustrated in Figure~1 by evenly distributing 100 radials over a 90 degree polar angle. For simplicity, dot 3 is empty and held deep in Coulomb blockade ($V_\mathrm{3}$=constant), such that the device can be viewed as a double dot underneath gates 1 and 2 that is capacitively coupled to a sensor dot under gate 4.  
As exemplified in Fig.~1d, ramps of control voltages $V_\mathrm{1}$ and $V_\mathrm{2}$ applied by a QDevil QDAC, are compensated by simultaneous ramps applied to $V_\mathrm{4}$, yielding 100 time stamps plotted in Fig.~1b. 

For our choice of reflectometry settings, $V_\mathrm{H}$ is high whenever dot 4 exchanges electrons with its reservoir, and low whenever dot 4 is in Coulomb blockade. Experimentally, this is accomplished by wirebonding the reflectometry tank circuit to gate 4 and by accumulating 8-9 electrons in dot 4 such that an enhanced reflectometry signal arises at the charge degeneracy between dot 4 and its reservoir~\cite{Ansaloni2020,Bohuslavskyi2020}.

\section{Appendix B: Active Learning Algorithm}
Generally, adding a linear equality constraint to the ground state of a convex polytope also leads to a convex polytope~\cite{nazarov2009quantum, Henk2017}. 
For the 2$\times$2 array, this implies that the boundary of a particular charge state (here of 1118) consists of a set of convex polytopes (3D manifolds embedded in 4D voltage space), for example the dot-4 facet (i.e. the degeneracy between 1118 and 1119).

Specifically, the algorithm starts without knowledge of any time stamps, executing an initial set of randomly oriented line searches emanating from a point within the dot-4 facet provided by the user. 
The position of this initial point can be anywhere within the charge state, as long as the sensor dot is set to the degeneracy with its reservoir and appropriately compensated~\cite{Bohuslavskyi2020}, which we intend to automate in future.  
For each resulting trigger event, the algorithm stores the radial distance ($r$) and direction, and calculates a confidence interval $(r^-,r^+)$, where $r^\pm = r\pm \delta/2$.  The value of $\delta$ is fixed by the user prior to the experiment, such that the radial interval contains the true boundary with high confidence, i.e. $\delta$ represents a tradeoff between line search precision and measurement time. For the results presented here, $\delta$ was set to 2 mV for a nominal ramp speed of 3 s per line search. Each search takes significantly less time since it is terminated and followed by the next search as soon as a trigger event occurs.

Following the initial set of measurements, a large-margin classifier estimates each plane of the polytope, defined as convex relaxations of a large-margin classification problem designed to produce solutions with as few facets as possible (illustrated on the left of Fig.~4a by a red triangle). 
This part of the algorithm is implemented using the cvxpy library~\cite{diamond2016cvxpy, agrawal2018rewriting}, specifically the ECOS solver for second-order cone problems~\cite{Domahidi2013ecos}. 
Halfspace intersections, convex hulls and their volumes are calculated using the Qhull library \cite{barber1996quickhull}. 

The full machine-learning task, however, extends beyond pure polytope estimation from measurements. This is because our goal is not to find the optimal polytope given a fixed set of points, but a polytope that accurately reflects the true polytope with all its facets (ground truth). Therefore, one cannot rely on an i.i.d. (independent and identically distributed random variables) dataset, but must concentrate line searches strategically in regions of high information gain. This is because facets with high operational value, such as inter-dot transitions used for Heisenberg spin exchange or quantum cellular automata, tend to occupy small regions in gate space, making them difficult to find by random sampling. After each iteration of the classifier, the algorithm therefore switches to an active-learning scheme that iteratively improves the polytope estimate by gathering new points to systematically disprove the previous estimate (green arrows in Fig.~4a).  
This is achieved by executing line searches directed through each side of the current estimate (to validate each side) and towards the corners of the estimate (to test for so-far undetected planes). 
If the new points disprove the current estimate, i.e. if the new confidence intervals $(r^+, r^-)$ lie fully inside or outside the estimate, the classifier estimates a new convex polytope based on the new points.


\begin{thebibliography}{50}%
	\makeatletter
	\providecommand \@ifxundefined [1]{%
		\@ifx{#1\undefined}
	}%
	\providecommand \@ifnum [1]{%
		\ifnum #1\expandafter \@firstoftwo
		\else \expandafter \@secondoftwo
		\fi
	}%
	\providecommand \@ifx [1]{%
		\ifx #1\expandafter \@firstoftwo
		\else \expandafter \@secondoftwo
		\fi
	}%
	\providecommand \natexlab [1]{#1}%
	\providecommand \enquote  [1]{``#1''}%
	\providecommand \bibnamefont  [1]{#1}%
	\providecommand \bibfnamefont [1]{#1}%
	\providecommand \citenamefont [1]{#1}%
	\providecommand \href@noop [0]{\@secondoftwo}%
	\providecommand \href [0]{\begingroup \@sanitize@url \@href}%
	\providecommand \@href[1]{\@@startlink{#1}\@@href}%
	\providecommand \@@href[1]{\endgroup#1\@@endlink}%
	\providecommand \@sanitize@url [0]{\catcode `\\12\catcode `\$12\catcode
		`\&12\catcode `\#12\catcode `\^12\catcode `\_12\catcode `\%12\relax}%
	\providecommand \@@startlink[1]{}%
	\providecommand \@@endlink[0]{}%
	\providecommand \url  [0]{\begingroup\@sanitize@url \@url }%
	\providecommand \@url [1]{\endgroup\@href {#1}{\urlprefix }}%
	\providecommand \urlprefix  [0]{URL }%
	\providecommand \Eprint [0]{\href }%
	\providecommand \doibase [0]{http://dx.doi.org/}%
	\providecommand \selectlanguage [0]{\@gobble}%
	\providecommand \bibinfo  [0]{\@secondoftwo}%
	\providecommand \bibfield  [0]{\@secondoftwo}%
	\providecommand \translation [1]{[#1]}%
	\providecommand \BibitemOpen [0]{}%
	\providecommand \bibitemStop [0]{}%
	\providecommand \bibitemNoStop [0]{.\EOS\space}%
	\providecommand \EOS [0]{\spacefactor3000\relax}%
	\providecommand \BibitemShut  [1]{\csname bibitem#1\endcsname}%
	\let\auto@bib@innerbib\@empty
	\bibitem [{\citenamefont {Yoneda}\ \emph {et~al.}(2018)\citenamefont {Yoneda},
		\citenamefont {Takeda}, \citenamefont {Otsuka}, \citenamefont {Nakajima},
		\citenamefont {Delbecq}, \citenamefont {Allison}, \citenamefont {Honda},
		\citenamefont {Kodera}, \citenamefont {Oda}, \citenamefont {Hoshi},
		\citenamefont {Usami}, \citenamefont {Itoh},\ and\ \citenamefont
		{Tarucha}}]{Yoneda2018}%
	\BibitemOpen
	\bibfield  {author} {\bibinfo {author} {\bibfnamefont {J.}~\bibnamefont
			{Yoneda}}, \bibinfo {author} {\bibfnamefont {K.}~\bibnamefont {Takeda}},
		\bibinfo {author} {\bibfnamefont {T.}~\bibnamefont {Otsuka}}, \bibinfo
		{author} {\bibfnamefont {T.}~\bibnamefont {Nakajima}}, \bibinfo {author}
		{\bibfnamefont {M.~R.}\ \bibnamefont {Delbecq}}, \bibinfo {author}
		{\bibfnamefont {G.}~\bibnamefont {Allison}}, \bibinfo {author} {\bibfnamefont
			{T.}~\bibnamefont {Honda}}, \bibinfo {author} {\bibfnamefont
			{T.}~\bibnamefont {Kodera}}, \bibinfo {author} {\bibfnamefont
			{S.}~\bibnamefont {Oda}}, \bibinfo {author} {\bibfnamefont {Y.}~\bibnamefont
			{Hoshi}}, \bibinfo {author} {\bibfnamefont {N.}~\bibnamefont {Usami}},
		\bibinfo {author} {\bibfnamefont {K.~M.}\ \bibnamefont {Itoh}}, \ and\
		\bibinfo {author} {\bibfnamefont {S.}~\bibnamefont {Tarucha}},\ }\bibfield
	{title} {\enquote {\bibinfo {title} {A quantum-dot spin qubit with coherence
				limited by charge noise and fidelity higher than 99.9\%},}\ }\href@noop {}
	{\bibfield  {journal} {\bibinfo  {journal} {Nature Nanotechnology}\ }\textbf
		{\bibinfo {volume} {13}},\ \bibinfo {pages} {102--106} (\bibinfo {year}
		{2018})}\BibitemShut {NoStop}%
	\bibitem [{\citenamefont {Muhonen}\ \emph {et~al.}(2014)\citenamefont
		{Muhonen}, \citenamefont {Dehollain}, \citenamefont {Laucht}, \citenamefont
		{Hudson}, \citenamefont {Kalra}, \citenamefont {Sekiguchi}, \citenamefont
		{Itoh}, \citenamefont {Jamieson}, \citenamefont {Mccallum}, \citenamefont
		{Dzurak},\ and\ \citenamefont {Morello}}]{Muhonen2014}%
	\BibitemOpen
	\bibfield  {author} {\bibinfo {author} {\bibfnamefont {J.~T.}\ \bibnamefont
			{Muhonen}}, \bibinfo {author} {\bibfnamefont {J.~P.}\ \bibnamefont
			{Dehollain}}, \bibinfo {author} {\bibfnamefont {A.}~\bibnamefont {Laucht}},
		\bibinfo {author} {\bibfnamefont {F.~E.}\ \bibnamefont {Hudson}}, \bibinfo
		{author} {\bibfnamefont {R.}~\bibnamefont {Kalra}}, \bibinfo {author}
		{\bibfnamefont {T.}~\bibnamefont {Sekiguchi}}, \bibinfo {author}
		{\bibfnamefont {K.~M.}\ \bibnamefont {Itoh}}, \bibinfo {author}
		{\bibfnamefont {D.~N.}\ \bibnamefont {Jamieson}}, \bibinfo {author}
		{\bibfnamefont {J.~C.}\ \bibnamefont {Mccallum}}, \bibinfo {author}
		{\bibfnamefont {A.~S.}\ \bibnamefont {Dzurak}}, \ and\ \bibinfo {author}
		{\bibfnamefont {A.}~\bibnamefont {Morello}},\ }\bibfield  {title} {\enquote
		{\bibinfo {title} {{Storing quantum information for 30 seconds in a
					nanoelectronic device}},}\ }\href {\doibase 10.1038/nnano.2014.211}
	{\bibfield  {journal} {\bibinfo  {journal} {Nature Nanotechnology}\ }\textbf
		{\bibinfo {volume} {9}},\ \bibinfo {pages} {986} (\bibinfo {year}
		{2014})}\BibitemShut {NoStop}%
	\bibitem [{\citenamefont {Watson}\ \emph {et~al.}(2018)\citenamefont {Watson},
		\citenamefont {Philips}, \citenamefont {Kawakami}, \citenamefont {Ward},
		\citenamefont {Scarlino}, \citenamefont {Veldhorst}, \citenamefont {Savage},
		\citenamefont {Lagally}, \citenamefont {Friesen}, \citenamefont
		{Coppersmith}, \citenamefont {Eriksson},\ and\ \citenamefont
		{Vandersypen}}]{Watson2018}%
	\BibitemOpen
	\bibfield  {author} {\bibinfo {author} {\bibfnamefont {T.~F.}\ \bibnamefont
			{Watson}}, \bibinfo {author} {\bibfnamefont {S.~G.~J.}\ \bibnamefont
			{Philips}}, \bibinfo {author} {\bibfnamefont {E.}~\bibnamefont {Kawakami}},
		\bibinfo {author} {\bibfnamefont {D.~R.}\ \bibnamefont {Ward}}, \bibinfo
		{author} {\bibfnamefont {P.}~\bibnamefont {Scarlino}}, \bibinfo {author}
		{\bibfnamefont {M.}~\bibnamefont {Veldhorst}}, \bibinfo {author}
		{\bibfnamefont {D.~E.}\ \bibnamefont {Savage}}, \bibinfo {author}
		{\bibfnamefont {M.~G.}\ \bibnamefont {Lagally}}, \bibinfo {author}
		{\bibfnamefont {M.}~\bibnamefont {Friesen}}, \bibinfo {author} {\bibfnamefont
			{S.~N.}\ \bibnamefont {Coppersmith}}, \bibinfo {author} {\bibfnamefont
			{M.~A.}\ \bibnamefont {Eriksson}}, \ and\ \bibinfo {author} {\bibfnamefont
			{L.~M.~K.}\ \bibnamefont {Vandersypen}},\ }\bibfield  {title} {\enquote
		{\bibinfo {title} {{A programmable two-qubit quantum processor in
					silicon}},}\ }\href {\doibase 10.1038/nature25766} {\bibfield  {journal}
		{\bibinfo  {journal} {Nature}\ }\textbf {\bibinfo {volume} {555}},\ \bibinfo
		{pages} {633--637} (\bibinfo {year} {2018})}\BibitemShut {NoStop}%
	\bibitem [{\citenamefont {Yang}\ \emph {et~al.}(2020)\citenamefont {Yang},
		\citenamefont {Leon}, \citenamefont {Hwang}, \citenamefont {Saraiva},
		\citenamefont {Tanttu}, \citenamefont {Huang}, \citenamefont
		{Camirand~Lemyre}, \citenamefont {Chan}, \citenamefont {Tan}, \citenamefont
		{Hudson}, \citenamefont {Itoh}, \citenamefont {Morello}, \citenamefont
		{Pioro-Ladri{\`e}re}, \citenamefont {Laucht},\ and\ \citenamefont
		{Dzurak}}]{Yang2020}%
	\BibitemOpen
	\bibfield  {author} {\bibinfo {author} {\bibfnamefont {C~H}\ \bibnamefont
			{Yang}}, \bibinfo {author} {\bibfnamefont {R~C~C}\ \bibnamefont {Leon}},
		\bibinfo {author} {\bibfnamefont {J~C~C}\ \bibnamefont {Hwang}}, \bibinfo
		{author} {\bibfnamefont {A}~\bibnamefont {Saraiva}}, \bibinfo {author}
		{\bibfnamefont {T}~\bibnamefont {Tanttu}}, \bibinfo {author} {\bibfnamefont
			{W}~\bibnamefont {Huang}}, \bibinfo {author} {\bibfnamefont {J}~\bibnamefont
			{Camirand~Lemyre}}, \bibinfo {author} {\bibfnamefont {K~W}\ \bibnamefont
			{Chan}}, \bibinfo {author} {\bibfnamefont {K~Y}\ \bibnamefont {Tan}},
		\bibinfo {author} {\bibfnamefont {F~E}\ \bibnamefont {Hudson}}, \bibinfo
		{author} {\bibfnamefont {K~M}\ \bibnamefont {Itoh}}, \bibinfo {author}
		{\bibfnamefont {A}~\bibnamefont {Morello}}, \bibinfo {author} {\bibfnamefont
			{M}~\bibnamefont {Pioro-Ladri{\`e}re}}, \bibinfo {author} {\bibfnamefont
			{A}~\bibnamefont {Laucht}}, \ and\ \bibinfo {author} {\bibfnamefont {A~S}\
			\bibnamefont {Dzurak}},\ }\bibfield  {title} {\enquote {\bibinfo {title}
			{{Operation of a silicon quantum processor unit cell above one kelvin}},}\
	}\href@noop {} {\bibfield  {journal} {\bibinfo  {journal} {Nature}\ }\textbf
		{\bibinfo {volume} {580}},\ \bibinfo {pages} {350--354} (\bibinfo {year}
		{2020})}\BibitemShut {NoStop}%
	\bibitem [{\citenamefont {Zajac}\ \emph {et~al.}(2016)\citenamefont {Zajac},
		\citenamefont {Hazard}, \citenamefont {Mi}, \citenamefont {Nielsen},
		\citenamefont {Petta},\ and\ \citenamefont {Petta}}]{Zajac2016}%
	\BibitemOpen
	\bibfield  {author} {\bibinfo {author} {\bibfnamefont {D~M}\ \bibnamefont
			{Zajac}}, \bibinfo {author} {\bibfnamefont {T~M}\ \bibnamefont {Hazard}},
		\bibinfo {author} {\bibfnamefont {X}~\bibnamefont {Mi}}, \bibinfo {author}
		{\bibfnamefont {E}~\bibnamefont {Nielsen}}, \bibinfo {author} {\bibfnamefont
			{J~R}\ \bibnamefont {Petta}}, \ and\ \bibinfo {author} {\bibfnamefont {J~R}\
			\bibnamefont {Petta}},\ }\bibfield  {title} {\enquote {\bibinfo {title}
			{{Scalable Gate Architecture for a One-Dimensional Array of Semiconductor
					Spin Qubits}},}\ }\href@noop {} {\bibfield  {journal} {\bibinfo  {journal}
			{Phys. Rev. Applied}\ }\textbf {\bibinfo {volume} {6}},\ \bibinfo {pages}
		{054013} (\bibinfo {year} {2016})}\BibitemShut {NoStop}%
	\bibitem [{\citenamefont {Volk}\ \emph
		{et~al.}(2019{\natexlab{a}})\citenamefont {Volk}, \citenamefont {Zwerver},
		\citenamefont {Mukhopadhyay}, \citenamefont {Eendebak}, \citenamefont {van
			Diepen}, \citenamefont {Dehollain}, \citenamefont {Hensgens}, \citenamefont
		{Fujita}, \citenamefont {Reichl}, \citenamefont {Wegscheider},\ and\
		\citenamefont {Vandersypen}}]{Volk2019}%
	\BibitemOpen
	\bibfield  {author} {\bibinfo {author} {\bibfnamefont {C}~\bibnamefont
			{Volk}}, \bibinfo {author} {\bibfnamefont {A~M~J}\ \bibnamefont {Zwerver}},
		\bibinfo {author} {\bibfnamefont {U}~\bibnamefont {Mukhopadhyay}}, \bibinfo
		{author} {\bibfnamefont {P~T}\ \bibnamefont {Eendebak}}, \bibinfo {author}
		{\bibfnamefont {C~J}\ \bibnamefont {van Diepen}}, \bibinfo {author}
		{\bibfnamefont {J~P}\ \bibnamefont {Dehollain}}, \bibinfo {author}
		{\bibfnamefont {T}~\bibnamefont {Hensgens}}, \bibinfo {author} {\bibfnamefont
			{T}~\bibnamefont {Fujita}}, \bibinfo {author} {\bibfnamefont {C}~\bibnamefont
			{Reichl}}, \bibinfo {author} {\bibfnamefont {W}~\bibnamefont {Wegscheider}},
		\ and\ \bibinfo {author} {\bibfnamefont {L~M~K}\ \bibnamefont
			{Vandersypen}},\ }\bibfield  {title} {\enquote {\bibinfo {title} {{Loading a
					quantum-dot based ``Qubyte'' register}},}\ }\href@noop {} {\bibfield
		{journal} {\bibinfo  {journal} {npj Quantum Information}\ }\textbf {\bibinfo
			{volume} {5}},\ \bibinfo {pages} {29} (\bibinfo {year}
		{2019}{\natexlab{a}})}\BibitemShut {NoStop}%
	\bibitem [{\citenamefont {Mills}\ \emph {et~al.}(2019)\citenamefont {Mills},
		\citenamefont {Zajac}, \citenamefont {Gullans}, \citenamefont {Schupp},
		\citenamefont {Hazard},\ and\ \citenamefont {Petta}}]{Mills2019}%
	\BibitemOpen
	\bibfield  {author} {\bibinfo {author} {\bibfnamefont {A.R.}\ \bibnamefont
			{Mills}}, \bibinfo {author} {\bibfnamefont {D.M.}\ \bibnamefont {Zajac}},
		\bibinfo {author} {\bibfnamefont {M.J.}\ \bibnamefont {Gullans}}, \bibinfo
		{author} {\bibfnamefont {F.J.}\ \bibnamefont {Schupp}}, \bibinfo {author}
		{\bibfnamefont {T.M.}\ \bibnamefont {Hazard}}, \ and\ \bibinfo {author}
		{\bibfnamefont {J.R.}\ \bibnamefont {Petta}},\ }\bibfield  {title} {\enquote
		{\bibinfo {title} {{Shuttling a single charge across a one-dimensional array
					of silicon quantum dots}},}\ }\href@noop {} {\bibfield  {journal} {\bibinfo
			{journal} {Nature Communications}\ }\textbf {\bibinfo {volume} {10}},\
		\bibinfo {pages} {1063} (\bibinfo {year} {2019})}\BibitemShut {NoStop}%
	\bibitem [{\citenamefont {Lawrie}\ \emph {et~al.}(2020)\citenamefont {Lawrie},
		\citenamefont {Eenink}, \citenamefont {Hendrickx}, \citenamefont {Boter},
		\citenamefont {Petit}, \citenamefont {Amitonov}, \citenamefont {Lodari},
		\citenamefont {Wuetz}, \citenamefont {Volk}, \citenamefont {Philips},
		\citenamefont {Droulers}, \citenamefont {Kalhor}, \citenamefont {van
			Riggelen}, \citenamefont {Brousse}, \citenamefont {Sammak}, \citenamefont
		{Vandersypen}, \citenamefont {Scappucci},\ and\ \citenamefont
		{Veldhorst}}]{Lawrie2019}%
	\BibitemOpen
	\bibfield  {author} {\bibinfo {author} {\bibfnamefont {W.~I.~L.}\
			\bibnamefont {Lawrie}}, \bibinfo {author} {\bibfnamefont {H.~G.~J.}\
			\bibnamefont {Eenink}}, \bibinfo {author} {\bibfnamefont {N.~W.}\
			\bibnamefont {Hendrickx}}, \bibinfo {author} {\bibfnamefont {J.~M.}\
			\bibnamefont {Boter}}, \bibinfo {author} {\bibfnamefont {L.}~\bibnamefont
			{Petit}}, \bibinfo {author} {\bibfnamefont {S.~V.}\ \bibnamefont {Amitonov}},
		\bibinfo {author} {\bibfnamefont {M.}~\bibnamefont {Lodari}}, \bibinfo
		{author} {\bibfnamefont {B.~Paquelet}\ \bibnamefont {Wuetz}}, \bibinfo
		{author} {\bibfnamefont {C.}~\bibnamefont {Volk}}, \bibinfo {author}
		{\bibfnamefont {S.}~\bibnamefont {Philips}}, \bibinfo {author} {\bibfnamefont
			{G.}~\bibnamefont {Droulers}}, \bibinfo {author} {\bibfnamefont
			{N.}~\bibnamefont {Kalhor}}, \bibinfo {author} {\bibfnamefont
			{F.}~\bibnamefont {van Riggelen}}, \bibinfo {author} {\bibfnamefont
			{D.}~\bibnamefont {Brousse}}, \bibinfo {author} {\bibfnamefont
			{A.}~\bibnamefont {Sammak}}, \bibinfo {author} {\bibfnamefont {L.~M.~K.}\
			\bibnamefont {Vandersypen}}, \bibinfo {author} {\bibfnamefont
			{G.}~\bibnamefont {Scappucci}}, \ and\ \bibinfo {author} {\bibfnamefont
			{M.}~\bibnamefont {Veldhorst}},\ }\bibfield  {title} {\enquote {\bibinfo
			{title} {{Quantum Dot Arrays in Silicon and Germanium}},}\ }\href@noop {}
	{\bibfield  {journal} {\bibinfo  {journal} {Applied Physics Letters}\
		}\textbf {\bibinfo {volume} {116}},\ \bibinfo {pages} {080501} (\bibinfo
		{year} {2020})}\BibitemShut {NoStop}%
	\bibitem [{\citenamefont {Dehollain}\ \emph {et~al.}(2020)\citenamefont
		{Dehollain}, \citenamefont {Mukhopadhyay}, \citenamefont {Michal},
		\citenamefont {Wang}, \citenamefont {Wunsch}, \citenamefont {Reichl},
		\citenamefont {Wegscheider}, \citenamefont {Rudner}, \citenamefont {Demler},\
		and\ \citenamefont {Vandersypen}}]{Dehollain2020}%
	\BibitemOpen
	\bibfield  {author} {\bibinfo {author} {\bibfnamefont {J~P}\ \bibnamefont
			{Dehollain}}, \bibinfo {author} {\bibfnamefont {U}~\bibnamefont
			{Mukhopadhyay}}, \bibinfo {author} {\bibfnamefont {V~P}\ \bibnamefont
			{Michal}}, \bibinfo {author} {\bibfnamefont {Y}~\bibnamefont {Wang}},
		\bibinfo {author} {\bibfnamefont {B}~\bibnamefont {Wunsch}}, \bibinfo
		{author} {\bibfnamefont {C}~\bibnamefont {Reichl}}, \bibinfo {author}
		{\bibfnamefont {W}~\bibnamefont {Wegscheider}}, \bibinfo {author}
		{\bibfnamefont {M~S}\ \bibnamefont {Rudner}}, \bibinfo {author}
		{\bibfnamefont {E}~\bibnamefont {Demler}}, \ and\ \bibinfo {author}
		{\bibfnamefont {L~M~K}\ \bibnamefont {Vandersypen}},\ }\bibfield  {title}
	{\enquote {\bibinfo {title} {{Nagaoka ferromagnetism observed in a quantum
					dot plaquette}},}\ }\href@noop {} {\bibfield  {journal} {\bibinfo  {journal}
			{Nature}\ }\textbf {\bibinfo {volume} {579}},\ \bibinfo {pages} {528--533}
		(\bibinfo {year} {2020})}\BibitemShut {NoStop}%
	\bibitem [{\citenamefont {Ansaloni}\ \emph {et~al.}(2020)\citenamefont
		{Ansaloni}, \citenamefont {Chatterjee}, \citenamefont {Bohuslavskyi},
		\citenamefont {Bertrand}, \citenamefont {Hutin}, \citenamefont {Vinet},\ and\
		\citenamefont {Kuemmeth}}]{Ansaloni2020}%
	\BibitemOpen
	\bibfield  {author} {\bibinfo {author} {\bibfnamefont {Fabio}\ \bibnamefont
			{Ansaloni}}, \bibinfo {author} {\bibfnamefont {Anasua}\ \bibnamefont
			{Chatterjee}}, \bibinfo {author} {\bibfnamefont {Heorhii}\ \bibnamefont
			{Bohuslavskyi}}, \bibinfo {author} {\bibfnamefont {Benoit}\ \bibnamefont
			{Bertrand}}, \bibinfo {author} {\bibfnamefont {Louis}\ \bibnamefont {Hutin}},
		\bibinfo {author} {\bibfnamefont {Maud}\ \bibnamefont {Vinet}}, \ and\
		\bibinfo {author} {\bibfnamefont {Ferdinand}\ \bibnamefont {Kuemmeth}},\
	}\bibfield  {title} {\enquote {\bibinfo {title} {{Single-electron operations
					in a foundry-fabricated array of quantum dots}},}\ }\href@noop {} {\bibfield
		{journal} {\bibinfo  {journal} {Nature Communications}\ }\textbf {\bibinfo
			{volume} {11}},\ \bibinfo {pages} {6399} (\bibinfo {year}
		{2020})}\BibitemShut {NoStop}%
	\bibitem [{\citenamefont {Mortemousque}\ \emph {et~al.}(2020)\citenamefont
		{Mortemousque}, \citenamefont {Chanrion}, \citenamefont {Jadot},
		\citenamefont {Flentje}, \citenamefont {Ludwig}, \citenamefont {Wieck},
		\citenamefont {Urdampilleta}, \citenamefont {Christopher},\ and\
		\citenamefont {Meunier}}]{Mortemousque2020}%
	\BibitemOpen
	\bibfield  {author} {\bibinfo {author} {\bibfnamefont {P.A.}\ \bibnamefont
			{Mortemousque}}, \bibinfo {author} {\bibfnamefont {E.}~\bibnamefont
			{Chanrion}}, \bibinfo {author} {\bibfnamefont {B.}~\bibnamefont {Jadot}},
		\bibinfo {author} {\bibfnamefont {H.}~\bibnamefont {Flentje}}, \bibinfo
		{author} {\bibfnamefont {A.}~\bibnamefont {Ludwig}}, \bibinfo {author}
		{\bibfnamefont {A.D.}\ \bibnamefont {Wieck}}, \bibinfo {author}
		{\bibfnamefont {M.}~\bibnamefont {Urdampilleta}}, \bibinfo {author}
		{\bibfnamefont {B.}~\bibnamefont {Christopher}}, \ and\ \bibinfo {author}
		{\bibfnamefont {T.}~\bibnamefont {Meunier}},\ }\bibfield  {title} {\enquote
		{\bibinfo {title} {{Coherent control of individual electron spins in a two
					dimensional array of quantum dots}},}\ }\href@noop {} {\bibfield  {journal}
		{\bibinfo  {journal} {Nature Nanotechnology}\ }\textbf {\bibinfo {volume}
			{16}},\ \bibinfo {pages} {296--301} (\bibinfo {year} {2020})}\BibitemShut
	{NoStop}%
	\bibitem [{\citenamefont {van Riggelen}\ \emph {et~al.}(2021)\citenamefont {van
			Riggelen}, \citenamefont {Hendrickx}, \citenamefont {Lawrie}, \citenamefont
		{Russ}, \citenamefont {Sammak}, \citenamefont {Scappucci},\ and\
		\citenamefont {Veldhorst}}]{vanRiggelen2021}%
	\BibitemOpen
	\bibfield  {author} {\bibinfo {author} {\bibfnamefont {F.}~\bibnamefont {van
				Riggelen}}, \bibinfo {author} {\bibfnamefont {N.~W.}\ \bibnamefont
			{Hendrickx}}, \bibinfo {author} {\bibfnamefont {W.~I.~L.}\ \bibnamefont
			{Lawrie}}, \bibinfo {author} {\bibfnamefont {M.}~\bibnamefont {Russ}},
		\bibinfo {author} {\bibfnamefont {A.}~\bibnamefont {Sammak}}, \bibinfo
		{author} {\bibfnamefont {G.}~\bibnamefont {Scappucci}}, \ and\ \bibinfo
		{author} {\bibfnamefont {M.}~\bibnamefont {Veldhorst}},\ }\bibfield  {title}
	{\enquote {\bibinfo {title} {A two-dimensional array of single-hole quantum
				dots},}\ }\href {\doibase 10.1063/5.0037330} {\bibfield  {journal} {\bibinfo
			{journal} {Applied Physics Letters}\ }\textbf {\bibinfo {volume} {118}},\
		\bibinfo {pages} {044002} (\bibinfo {year} {2021})}\BibitemShut {NoStop}%
	\bibitem [{\citenamefont {Fedele}\ \emph {et~al.}(2021)\citenamefont {Fedele},
		\citenamefont {Chatterjee}, \citenamefont {Fallahi}, \citenamefont {Gardner},
		\citenamefont {Manfra},\ and\ \citenamefont {Kuemmeth}}]{Fedele2021}%
	\BibitemOpen
	\bibfield  {author} {\bibinfo {author} {\bibfnamefont {Federico}\
			\bibnamefont {Fedele}}, \bibinfo {author} {\bibfnamefont {Anasua}\
			\bibnamefont {Chatterjee}}, \bibinfo {author} {\bibfnamefont {Saeed}\
			\bibnamefont {Fallahi}}, \bibinfo {author} {\bibfnamefont {Geoffrey~C.}\
			\bibnamefont {Gardner}}, \bibinfo {author} {\bibfnamefont {Michael~J.}\
			\bibnamefont {Manfra}}, \ and\ \bibinfo {author} {\bibfnamefont {Ferdinand}\
			\bibnamefont {Kuemmeth}},\ }\bibfield  {title} {\enquote {\bibinfo {title}
			{Simultaneous operations in a two-dimensional array of singlet-triplet
				qubits},}\ }\href@noop {} {\bibfield  {journal} {\bibinfo  {journal} {PRX
				Quantum}\ }\textbf {\bibinfo {volume} {2}},\ \bibinfo {pages} {040306}
		(\bibinfo {year} {2021})}\BibitemShut {NoStop}%
	\bibitem [{\citenamefont {Zwolak}\ \emph {et~al.}(2020)\citenamefont {Zwolak},
		\citenamefont {McJunkin}, \citenamefont {Kalantre}, \citenamefont {Dodson},
		\citenamefont {MacQuarrie}, \citenamefont {Savage}, \citenamefont {Lagally},
		\citenamefont {Coppersmith}, \citenamefont {Eriksson},\ and\ \citenamefont
		{Taylor}}]{Zwolak2020}%
	\BibitemOpen
	\bibfield  {author} {\bibinfo {author} {\bibfnamefont {Justyna~P.}\
			\bibnamefont {Zwolak}}, \bibinfo {author} {\bibfnamefont {Thomas}\
			\bibnamefont {McJunkin}}, \bibinfo {author} {\bibfnamefont {Sandesh~S.}\
			\bibnamefont {Kalantre}}, \bibinfo {author} {\bibfnamefont {J.P.}\
			\bibnamefont {Dodson}}, \bibinfo {author} {\bibfnamefont {E.R.}\ \bibnamefont
			{MacQuarrie}}, \bibinfo {author} {\bibfnamefont {D.E.}\ \bibnamefont
			{Savage}}, \bibinfo {author} {\bibfnamefont {M.G.}\ \bibnamefont {Lagally}},
		\bibinfo {author} {\bibfnamefont {S.N.}\ \bibnamefont {Coppersmith}},
		\bibinfo {author} {\bibfnamefont {Mark~A.}\ \bibnamefont {Eriksson}}, \ and\
		\bibinfo {author} {\bibfnamefont {Jacob~M.}\ \bibnamefont {Taylor}},\
	}\bibfield  {title} {\enquote {\bibinfo {title} {Autotuning of double-dot
				devices in situ with machine learning},}\ }\href@noop {} {\bibfield
		{journal} {\bibinfo  {journal} {Phys. Rev. Applied}\ }\textbf {\bibinfo
			{volume} {13}},\ \bibinfo {pages} {034075} (\bibinfo {year}
		{2020})}\BibitemShut {NoStop}%
	\bibitem [{\citenamefont {Zwolak}\ \emph {et~al.}(2021)\citenamefont {Zwolak},
		\citenamefont {McJunkin}, \citenamefont {Kalantre}, \citenamefont {Neyens},
		\citenamefont {MacQuarrie}, \citenamefont {Eriksson},\ and\ \citenamefont
		{Taylor}}]{Zwolak2021}%
	\BibitemOpen
	\bibfield  {author} {\bibinfo {author} {\bibfnamefont {Justyna~P.}\
			\bibnamefont {Zwolak}}, \bibinfo {author} {\bibfnamefont {Thomas}\
			\bibnamefont {McJunkin}}, \bibinfo {author} {\bibfnamefont {Sandesh~S.}\
			\bibnamefont {Kalantre}}, \bibinfo {author} {\bibfnamefont {Samuel~F.}\
			\bibnamefont {Neyens}}, \bibinfo {author} {\bibfnamefont {E.R.}\ \bibnamefont
			{MacQuarrie}}, \bibinfo {author} {\bibfnamefont {Mark~A.}\ \bibnamefont
			{Eriksson}}, \ and\ \bibinfo {author} {\bibfnamefont {Jacob~M.}\ \bibnamefont
			{Taylor}},\ }\bibfield  {title} {\enquote {\bibinfo {title} {Ray-based
				framework for state identification in quantum dot devices},}\ }\href
	{\doibase 10.1103/PRXQuantum.2.020335} {\bibfield  {journal} {\bibinfo
			{journal} {PRX Quantum}\ }\textbf {\bibinfo {volume} {2}},\ \bibinfo {pages}
		{020335} (\bibinfo {year} {2021})}\BibitemShut {NoStop}%
	\bibitem [{\citenamefont {Moon}\ \emph {et~al.}(2020)\citenamefont {Moon},
		\citenamefont {Lennon}, \citenamefont {Kirkpatrick}, \citenamefont {van
			Esbroeck}, \citenamefont {Camenzind}, \citenamefont {Yu}, \citenamefont
		{Vigneau}, \citenamefont {Zumb{\"u}hl}, \citenamefont {Briggs}, \citenamefont
		{Osborne}, \citenamefont {Sejdinovic}, \citenamefont {Laird},\ and\
		\citenamefont {Ares}}]{Moon2020}%
	\BibitemOpen
	\bibfield  {author} {\bibinfo {author} {\bibfnamefont {H}~\bibnamefont
			{Moon}}, \bibinfo {author} {\bibfnamefont {D~T}\ \bibnamefont {Lennon}},
		\bibinfo {author} {\bibfnamefont {J}~\bibnamefont {Kirkpatrick}}, \bibinfo
		{author} {\bibfnamefont {N~M}\ \bibnamefont {van Esbroeck}}, \bibinfo
		{author} {\bibfnamefont {L~C}\ \bibnamefont {Camenzind}}, \bibinfo {author}
		{\bibfnamefont {Liuqi}\ \bibnamefont {Yu}}, \bibinfo {author} {\bibfnamefont
			{F}~\bibnamefont {Vigneau}}, \bibinfo {author} {\bibfnamefont {D~M}\
			\bibnamefont {Zumb{\"u}hl}}, \bibinfo {author} {\bibfnamefont {G~A~D}\
			\bibnamefont {Briggs}}, \bibinfo {author} {\bibfnamefont {M~A}\ \bibnamefont
			{Osborne}}, \bibinfo {author} {\bibfnamefont {D}~\bibnamefont {Sejdinovic}},
		\bibinfo {author} {\bibfnamefont {E~A}\ \bibnamefont {Laird}}, \ and\
		\bibinfo {author} {\bibfnamefont {N}~\bibnamefont {Ares}},\ }\bibfield
	{title} {\enquote {\bibinfo {title} {{Machine learning enables completely
					automatic tuning of a quantum device faster than human experts}},}\
	}\href@noop {} {\bibfield  {journal} {\bibinfo  {journal} {Nature
				Communications}\ }\textbf {\bibinfo {volume} {11}},\ \bibinfo {pages} {4161}
		(\bibinfo {year} {2020})}\BibitemShut {NoStop}%
	\bibitem [{\citenamefont {Lennon}\ \emph {et~al.}(2019)\citenamefont {Lennon},
		\citenamefont {Moon}, \citenamefont {Camenzind}, \citenamefont {Yu},
		\citenamefont {Zumb{\"u}hl}, \citenamefont {Briggs}, \citenamefont {Osborne},
		\citenamefont {Laird},\ and\ \citenamefont {Ares}}]{Lennon2019}%
	\BibitemOpen
	\bibfield  {author} {\bibinfo {author} {\bibfnamefont {D~T}\ \bibnamefont
			{Lennon}}, \bibinfo {author} {\bibfnamefont {H}~\bibnamefont {Moon}},
		\bibinfo {author} {\bibfnamefont {L~C}\ \bibnamefont {Camenzind}}, \bibinfo
		{author} {\bibfnamefont {Liuqi}\ \bibnamefont {Yu}}, \bibinfo {author}
		{\bibfnamefont {D~M}\ \bibnamefont {Zumb{\"u}hl}}, \bibinfo {author}
		{\bibfnamefont {G~A~D}\ \bibnamefont {Briggs}}, \bibinfo {author}
		{\bibfnamefont {M~A}\ \bibnamefont {Osborne}}, \bibinfo {author}
		{\bibfnamefont {E~A}\ \bibnamefont {Laird}}, \ and\ \bibinfo {author}
		{\bibfnamefont {N}~\bibnamefont {Ares}},\ }\bibfield  {title} {\enquote
		{\bibinfo {title} {{Efficiently measuring a quantum device using machine
					learning}},}\ }\href@noop {} {\bibfield  {journal} {\bibinfo  {journal} {npj
				Quantum Information}\ }\textbf {\bibinfo {volume} {5}},\ \bibinfo {pages}
		{79} (\bibinfo {year} {2019})}\BibitemShut {NoStop}%
	\bibitem [{\citenamefont {Kalantre}\ \emph {et~al.}(2019)\citenamefont
		{Kalantre}, \citenamefont {Zwolak}, \citenamefont {Ragole}, \citenamefont
		{Wu}, \citenamefont {Zimmerman}, \citenamefont {Stewart},\ and\ \citenamefont
		{Taylor}}]{Kalantre2019}%
	\BibitemOpen
	\bibfield  {author} {\bibinfo {author} {\bibfnamefont {Sandesh~S}\
			\bibnamefont {Kalantre}}, \bibinfo {author} {\bibfnamefont {Justyna~P}\
			\bibnamefont {Zwolak}}, \bibinfo {author} {\bibfnamefont {Stephen}\
			\bibnamefont {Ragole}}, \bibinfo {author} {\bibfnamefont {Xingyao}\
			\bibnamefont {Wu}}, \bibinfo {author} {\bibfnamefont {Neil~M}\ \bibnamefont
			{Zimmerman}}, \bibinfo {author} {\bibfnamefont {Michael~D}\ \bibnamefont
			{Stewart}}, \ and\ \bibinfo {author} {\bibfnamefont {Jacob~M}\ \bibnamefont
			{Taylor}},\ }\bibfield  {title} {\enquote {\bibinfo {title} {Machine learning
				techniques for state recognition and auto-tuning in quantum dots},}\
	}\href@noop {} {\bibfield  {journal} {\bibinfo  {journal} {npj Quantum
				Information}\ }\textbf {\bibinfo {volume} {5}},\ \bibinfo {pages} {1--10}
		(\bibinfo {year} {2019})}\BibitemShut {NoStop}%
	\bibitem [{\citenamefont {Botzem}\ \emph {et~al.}(2018)\citenamefont {Botzem},
		\citenamefont {Shulman}, \citenamefont {Foletti}, \citenamefont {Harvey},
		\citenamefont {Dial}, \citenamefont {Bethke}, \citenamefont {Cerfontaine},
		\citenamefont {McNeil}, \citenamefont {Mahalu}, \citenamefont {Umansky} \emph
		{et~al.}}]{Botzem2018}%
	\BibitemOpen
	\bibfield  {author} {\bibinfo {author} {\bibfnamefont {Tim}\ \bibnamefont
			{Botzem}}, \bibinfo {author} {\bibfnamefont {Michael~D}\ \bibnamefont
			{Shulman}}, \bibinfo {author} {\bibfnamefont {Sandra}\ \bibnamefont
			{Foletti}}, \bibinfo {author} {\bibfnamefont {Shannon~P}\ \bibnamefont
			{Harvey}}, \bibinfo {author} {\bibfnamefont {Oliver~E}\ \bibnamefont {Dial}},
		\bibinfo {author} {\bibfnamefont {Patrick}\ \bibnamefont {Bethke}}, \bibinfo
		{author} {\bibfnamefont {Pascal}\ \bibnamefont {Cerfontaine}}, \bibinfo
		{author} {\bibfnamefont {Robert~PG}\ \bibnamefont {McNeil}}, \bibinfo
		{author} {\bibfnamefont {Diana}\ \bibnamefont {Mahalu}}, \bibinfo {author}
		{\bibfnamefont {Vladimir}\ \bibnamefont {Umansky}},  \emph {et~al.},\
	}\bibfield  {title} {\enquote {\bibinfo {title} {Tuning methods for
				semiconductor spin qubits},}\ }\href@noop {} {\bibfield  {journal} {\bibinfo
			{journal} {Physical Review Applied}\ }\textbf {\bibinfo {volume} {10}},\
		\bibinfo {pages} {054026} (\bibinfo {year} {2018})}\BibitemShut {NoStop}%
	\bibitem [{\citenamefont {Darulov\'a}\ \emph {et~al.}(2020)\citenamefont
		{Darulov\'a}, \citenamefont {Pauka}, \citenamefont {Wiebe}, \citenamefont
		{Chan}, \citenamefont {Gardener}, \citenamefont {Manfra}, \citenamefont
		{Cassidy},\ and\ \citenamefont {Troyer}}]{Darulova2020}%
	\BibitemOpen
	\bibfield  {author} {\bibinfo {author} {\bibfnamefont {J.}~\bibnamefont
			{Darulov\'a}}, \bibinfo {author} {\bibfnamefont {S.J.}\ \bibnamefont
			{Pauka}}, \bibinfo {author} {\bibfnamefont {N.}~\bibnamefont {Wiebe}},
		\bibinfo {author} {\bibfnamefont {K.W.}\ \bibnamefont {Chan}}, \bibinfo
		{author} {\bibfnamefont {G.C}\ \bibnamefont {Gardener}}, \bibinfo {author}
		{\bibfnamefont {M.J.}\ \bibnamefont {Manfra}}, \bibinfo {author}
		{\bibfnamefont {M.C.}\ \bibnamefont {Cassidy}}, \ and\ \bibinfo {author}
		{\bibfnamefont {M.}~\bibnamefont {Troyer}},\ }\bibfield  {title} {\enquote
		{\bibinfo {title} {Autonomous tuning and charge-state detection of
				gate-defined quantum dots},}\ }\href {\doibase
		10.1103/PhysRevApplied.13.054005} {\bibfield  {journal} {\bibinfo  {journal}
			{Phys. Rev. Applied}\ }\textbf {\bibinfo {volume} {13}},\ \bibinfo {pages}
		{054005} (\bibinfo {year} {2020})}\BibitemShut {NoStop}%
	\bibitem [{\citenamefont {Ziegler}\ \emph {et~al.}(2022)\citenamefont
		{Ziegler}, \citenamefont {Luthi}, \citenamefont {Ramsey}, \citenamefont
		{Borjans}, \citenamefont {Zheng},\ and\ \citenamefont
		{Zwolak}}]{Ziegler2022}%
	\BibitemOpen
	\bibfield  {author} {\bibinfo {author} {\bibfnamefont {J}~\bibnamefont
			{Ziegler}}, \bibinfo {author} {\bibfnamefont {F}~\bibnamefont {Luthi}},
		\bibinfo {author} {\bibfnamefont {M}~\bibnamefont {Ramsey}}, \bibinfo
		{author} {\bibfnamefont {F}~\bibnamefont {Borjans}}, \bibinfo {author}
		{\bibfnamefont {G}~\bibnamefont {Zheng}}, \ and\ \bibinfo {author}
		{\bibfnamefont {J.~P.}\ \bibnamefont {Zwolak}},\ }\bibfield  {title}
	{\enquote {\bibinfo {title} {Tuning arrays with rays: Physics-informed tuning
				of quantum dot charge states},}\ }\href@noop {} {\bibfield  {journal}
		{\bibinfo  {journal} {arXiv preprint arXiv:2209.03837}\ } (\bibinfo {year}
		{2022})}\BibitemShut {NoStop}%
	\bibitem [{\citenamefont {Chatterjee}\ \emph {et~al.}(2021)\citenamefont
		{Chatterjee}, \citenamefont {Stevenson}, \citenamefont {De~Franceschi},
		\citenamefont {Morello}, \citenamefont {de~Leon},\ and\ \citenamefont
		{Kuemmeth}}]{Chatterjee2021}%
	\BibitemOpen
	\bibfield  {author} {\bibinfo {author} {\bibfnamefont {A}~\bibnamefont
			{Chatterjee}}, \bibinfo {author} {\bibfnamefont {P}~\bibnamefont
			{Stevenson}}, \bibinfo {author} {\bibfnamefont {S}~\bibnamefont
			{De~Franceschi}}, \bibinfo {author} {\bibfnamefont {A}~\bibnamefont
			{Morello}}, \bibinfo {author} {\bibfnamefont {N~P}\ \bibnamefont {de~Leon}},
		\ and\ \bibinfo {author} {\bibfnamefont {F}~\bibnamefont {Kuemmeth}},\
	}\bibfield  {title} {\enquote {\bibinfo {title} {Semiconductor qubits in
				practice},}\ }\href@noop {} {\bibfield  {journal} {\bibinfo  {journal}
			{Nature Reviews Physics}\ }\textbf {\bibinfo {volume} {3}},\ \bibinfo {pages}
		{157--177} (\bibinfo {year} {2021})}\BibitemShut {NoStop}%
	\bibitem [{\citenamefont {Heinz}\ and\ \citenamefont
		{Burkard}(2021)}]{Heinz2021}%
	\BibitemOpen
	\bibfield  {author} {\bibinfo {author} {\bibfnamefont {Irina}\ \bibnamefont
			{Heinz}}\ and\ \bibinfo {author} {\bibfnamefont {Guido}\ \bibnamefont
			{Burkard}},\ }\bibfield  {title} {\enquote {\bibinfo {title} {Crosstalk
				analysis for single-qubit and two-qubit gates in spin qubit arrays},}\
	}\href@noop {} {\bibfield  {journal} {\bibinfo  {journal} {Phys. Rev. B}\
		}\textbf {\bibinfo {volume} {104}},\ \bibinfo {pages} {045420} (\bibinfo
		{year} {2021})}\BibitemShut {NoStop}%
	\bibitem [{\citenamefont {Bohuslavskyi}\ \emph {et~al.}(2020)\citenamefont
		{Bohuslavskyi}, \citenamefont {Ansaloni}, \citenamefont {Chatterjee},
		\citenamefont {Fedele}, \citenamefont {Rasmussen}, \citenamefont {Brovang},
		\citenamefont {Li}, \citenamefont {Hutin}, \citenamefont {Venitucci},
		\citenamefont {Bertrand}, \citenamefont {Vinet}, \citenamefont {Niquet},\
		and\ \citenamefont {Kuemmeth}}]{Bohuslavskyi2020}%
	\BibitemOpen
	\bibfield  {author} {\bibinfo {author} {\bibfnamefont {Heorhii}\ \bibnamefont
			{Bohuslavskyi}}, \bibinfo {author} {\bibfnamefont {Fabio}\ \bibnamefont
			{Ansaloni}}, \bibinfo {author} {\bibfnamefont {Anasua}\ \bibnamefont
			{Chatterjee}}, \bibinfo {author} {\bibfnamefont {Federico}\ \bibnamefont
			{Fedele}}, \bibinfo {author} {\bibfnamefont {Torbjørn}\ \bibnamefont
			{Rasmussen}}, \bibinfo {author} {\bibfnamefont {Bertram}\ \bibnamefont
			{Brovang}}, \bibinfo {author} {\bibfnamefont {Jing}\ \bibnamefont {Li}},
		\bibinfo {author} {\bibfnamefont {Louis}\ \bibnamefont {Hutin}}, \bibinfo
		{author} {\bibfnamefont {Benjamin}\ \bibnamefont {Venitucci}}, \bibinfo
		{author} {\bibfnamefont {Benoit}\ \bibnamefont {Bertrand}}, \bibinfo {author}
		{\bibfnamefont {Maud}\ \bibnamefont {Vinet}}, \bibinfo {author}
		{\bibfnamefont {Yann-Michel}\ \bibnamefont {Niquet}}, \ and\ \bibinfo
		{author} {\bibfnamefont {Ferdinand}\ \bibnamefont {Kuemmeth}},\ }\bibfield
	{title} {\enquote {\bibinfo {title} {Reflectometry of charge transitions in a
				silicon quadruple dot},}\ }\href@noop {} {\bibfield  {journal} {\bibinfo
			{journal} {arXiv preprint arXiv:2012.04791}\ } (\bibinfo {year}
		{2020})}\BibitemShut {NoStop}%
	\bibitem [{\citenamefont {Volk}\ \emph
		{et~al.}(2019{\natexlab{b}})\citenamefont {Volk}, \citenamefont {Chatterjee},
		\citenamefont {Ansaloni}, \citenamefont {Marcus},\ and\ \citenamefont
		{Kuemmeth}}]{Volk2019a}%
	\BibitemOpen
	\bibfield  {author} {\bibinfo {author} {\bibfnamefont {C}~\bibnamefont
			{Volk}}, \bibinfo {author} {\bibfnamefont {A}~\bibnamefont {Chatterjee}},
		\bibinfo {author} {\bibfnamefont {F}~\bibnamefont {Ansaloni}}, \bibinfo
		{author} {\bibfnamefont {C~M}\ \bibnamefont {Marcus}}, \ and\ \bibinfo
		{author} {\bibfnamefont {F}~\bibnamefont {Kuemmeth}},\ }\bibfield  {title}
	{\enquote {\bibinfo {title} {Fast charge sensing of {Si/SiGe} quantum dots
				via a high-frequency accumulation gate},}\ }\href@noop {} {\bibfield
		{journal} {\bibinfo  {journal} {Nano Letters}\ }\textbf {\bibinfo {volume}
			{19}},\ \bibinfo {pages} {5628--5633} (\bibinfo {year}
		{2019}{\natexlab{b}})}\BibitemShut {NoStop}%
	\bibitem [{QDe()}]{QDevil}%
	\BibitemOpen
	\href@noop {} {}\Eprint {http://arxiv.org/abs/Electronic access:
		https://www.qdevil.com} {Electronic access: https://www.qdevil.com}
	\BibitemShut {NoStop}%
	\bibitem [{\citenamefont {van~der Wiel}\ \emph {et~al.}(2002)\citenamefont
		{van~der Wiel}, \citenamefont {De~Franceschi}, \citenamefont {Elzerman},
		\citenamefont {Fujisawa}, \citenamefont {Tarucha},\ and\ \citenamefont
		{Kouwenhoven}}]{vanderWiel2002}%
	\BibitemOpen
	\bibfield  {author} {\bibinfo {author} {\bibfnamefont {W.~G.}\ \bibnamefont
			{van~der Wiel}}, \bibinfo {author} {\bibfnamefont {S.}~\bibnamefont
			{De~Franceschi}}, \bibinfo {author} {\bibfnamefont {J.~M.}\ \bibnamefont
			{Elzerman}}, \bibinfo {author} {\bibfnamefont {T.}~\bibnamefont {Fujisawa}},
		\bibinfo {author} {\bibfnamefont {S.}~\bibnamefont {Tarucha}}, \ and\
		\bibinfo {author} {\bibfnamefont {L.~P.}\ \bibnamefont {Kouwenhoven}},\
	}\href@noop {} {\bibfield  {journal} {\bibinfo  {journal} {Reviews of Modern
				Physics}\ }\textbf {\bibinfo {volume} {75}} (\bibinfo {year}
		{2002})}\BibitemShut {NoStop}%
	\bibitem [{\citenamefont {Medford}\ \emph {et~al.}(2013)\citenamefont
		{Medford}, \citenamefont {Beil}, \citenamefont {Taylor}, \citenamefont
		{Bartlett}, \citenamefont {Doherty}, \citenamefont {Rashba}, \citenamefont
		{Divincenzo}, \citenamefont {Lu}, \citenamefont {Gossard},\ and\
		\citenamefont {Marcus}}]{Medford2013}%
	\BibitemOpen
	\bibfield  {author} {\bibinfo {author} {\bibfnamefont {J.}~\bibnamefont
			{Medford}}, \bibinfo {author} {\bibfnamefont {J.}~\bibnamefont {Beil}},
		\bibinfo {author} {\bibfnamefont {J.~M.}\ \bibnamefont {Taylor}}, \bibinfo
		{author} {\bibfnamefont {S.~D.}\ \bibnamefont {Bartlett}}, \bibinfo {author}
		{\bibfnamefont {A.~C.}\ \bibnamefont {Doherty}}, \bibinfo {author}
		{\bibfnamefont {E.~I.}\ \bibnamefont {Rashba}}, \bibinfo {author}
		{\bibfnamefont {D.~P.}\ \bibnamefont {Divincenzo}}, \bibinfo {author}
		{\bibfnamefont {H.}~\bibnamefont {Lu}}, \bibinfo {author} {\bibfnamefont
			{A.~C.}\ \bibnamefont {Gossard}}, \ and\ \bibinfo {author} {\bibfnamefont
			{C.~M.}\ \bibnamefont {Marcus}},\ }\bibfield  {title} {\enquote {\bibinfo
			{title} {Self-consistent measurement and state tomography of an exchange-only
				spin qubit},}\ }\href@noop {} {\bibfield  {journal} {\bibinfo  {journal}
			{Nature Nanotechnology}\ }\textbf {\bibinfo {volume} {8}},\ \bibinfo {pages}
		{654--659} (\bibinfo {year} {2013})}\BibitemShut {NoStop}%
	\bibitem [{\citenamefont {Eng}\ \emph {et~al.}(2015)\citenamefont {Eng},
		\citenamefont {Ladd}, \citenamefont {Smith}, \citenamefont {Borselli},
		\citenamefont {Kiselev}, \citenamefont {Fong}, \citenamefont {Holabird},
		\citenamefont {Hazard}, \citenamefont {Huang}, \citenamefont {Deelman},
		\citenamefont {Milosavljevic}, \citenamefont {Schmitz}, \citenamefont {Ross},
		\citenamefont {Gyure},\ and\ \citenamefont {Hunter}}]{Eng2015}%
	\BibitemOpen
	\bibfield  {author} {\bibinfo {author} {\bibfnamefont {K.}~\bibnamefont
			{Eng}}, \bibinfo {author} {\bibfnamefont {Thaddeus~D}\ \bibnamefont {Ladd}},
		\bibinfo {author} {\bibfnamefont {A.}~\bibnamefont {Smith}}, \bibinfo
		{author} {\bibfnamefont {M.~G.}\ \bibnamefont {Borselli}}, \bibinfo {author}
		{\bibfnamefont {Andrey~A.}\ \bibnamefont {Kiselev}}, \bibinfo {author}
		{\bibfnamefont {B.~H.}\ \bibnamefont {Fong}}, \bibinfo {author}
		{\bibfnamefont {K.~S.}\ \bibnamefont {Holabird}}, \bibinfo {author}
		{\bibfnamefont {T.~M.}\ \bibnamefont {Hazard}}, \bibinfo {author}
		{\bibfnamefont {B.}~\bibnamefont {Huang}}, \bibinfo {author} {\bibfnamefont
			{P.~W.}\ \bibnamefont {Deelman}}, \bibinfo {author} {\bibfnamefont
			{I.}~\bibnamefont {Milosavljevic}}, \bibinfo {author} {\bibfnamefont {A.~E.}\
			\bibnamefont {Schmitz}}, \bibinfo {author} {\bibfnamefont {R.~S.}\
			\bibnamefont {Ross}}, \bibinfo {author} {\bibfnamefont {M.~F.}\ \bibnamefont
			{Gyure}}, \ and\ \bibinfo {author} {\bibfnamefont {A.~T.}\ \bibnamefont
			{Hunter}},\ }\bibfield  {title} {\enquote {\bibinfo {title} {{Isotopically
					enhanced triple-quantum-dot qubit}},}\ }\href@noop {} {\bibfield  {journal}
		{\bibinfo  {journal} {Science Advances}\ }\textbf {\bibinfo {volume} {1}},\
		\bibinfo {pages} {1500214} (\bibinfo {year} {2015})}\BibitemShut {NoStop}%
	\bibitem [{\citenamefont {Malinowski}\ \emph {et~al.}(2017)\citenamefont
		{Malinowski}, \citenamefont {Martins}, \citenamefont {Nissen}, \citenamefont
		{Fallahi}, \citenamefont {Gardner}, \citenamefont {Manfra}, \citenamefont
		{Marcus},\ and\ \citenamefont {Kuemmeth}}]{Malinowski2017b}%
	\BibitemOpen
	\bibfield  {author} {\bibinfo {author} {\bibfnamefont {Filip~K.}\
			\bibnamefont {Malinowski}}, \bibinfo {author} {\bibfnamefont {Frederico}\
			\bibnamefont {Martins}}, \bibinfo {author} {\bibfnamefont {Peter~D.}\
			\bibnamefont {Nissen}}, \bibinfo {author} {\bibfnamefont {Saeed}\
			\bibnamefont {Fallahi}}, \bibinfo {author} {\bibfnamefont {Geoffrey~C.}\
			\bibnamefont {Gardner}}, \bibinfo {author} {\bibfnamefont {Michael~J.}\
			\bibnamefont {Manfra}}, \bibinfo {author} {\bibfnamefont {Charles~M.}\
			\bibnamefont {Marcus}}, \ and\ \bibinfo {author} {\bibfnamefont {Ferdinand}\
			\bibnamefont {Kuemmeth}},\ }\bibfield  {title} {\enquote {\bibinfo {title}
			{{Symmetric operation of the resonant exchange qubit}},}\ }\href@noop {}
	{\bibfield  {journal} {\bibinfo  {journal} {Physical Review B}\ }\textbf
		{\bibinfo {volume} {96}},\ \bibinfo {pages} {045443} (\bibinfo {year}
		{2017})}\BibitemShut {NoStop}%
	\bibitem [{\citenamefont {Granger}\ \emph {et~al.}(2010)\citenamefont
		{Granger}, \citenamefont {Gaudreau}, \citenamefont {Kam}, \citenamefont
		{Pioro-Ladri\`ere}, \citenamefont {Studenikin}, \citenamefont {Wasilewski},
		\citenamefont {Zawadzki},\ and\ \citenamefont {Sachrajda}}]{Granger2010}%
	\BibitemOpen
	\bibfield  {author} {\bibinfo {author} {\bibfnamefont {G.}~\bibnamefont
			{Granger}}, \bibinfo {author} {\bibfnamefont {L.}~\bibnamefont {Gaudreau}},
		\bibinfo {author} {\bibfnamefont {A.}~\bibnamefont {Kam}}, \bibinfo {author}
		{\bibfnamefont {M.}~\bibnamefont {Pioro-Ladri\`ere}}, \bibinfo {author}
		{\bibfnamefont {S.~A.}\ \bibnamefont {Studenikin}}, \bibinfo {author}
		{\bibfnamefont {Z.~R.}\ \bibnamefont {Wasilewski}}, \bibinfo {author}
		{\bibfnamefont {P.}~\bibnamefont {Zawadzki}}, \ and\ \bibinfo {author}
		{\bibfnamefont {A.~S.}\ \bibnamefont {Sachrajda}},\ }\bibfield  {title}
	{\enquote {\bibinfo {title} {Three-dimensional transport diagram of a triple
				quantum dot},}\ }\href {\doibase 10.1103/PhysRevB.82.075304} {\bibfield
		{journal} {\bibinfo  {journal} {Phys. Rev. B}\ }\textbf {\bibinfo {volume}
			{82}},\ \bibinfo {pages} {075304} (\bibinfo {year} {2010})}\BibitemShut
	{NoStop}%
	\bibitem [{\citenamefont {Nazarov}\ and\ \citenamefont
		{Blanter}(2009)}]{nazarov2009quantum}%
	\BibitemOpen
	\bibfield  {author} {\bibinfo {author} {\bibfnamefont {Yuli~V}\ \bibnamefont
			{Nazarov}}\ and\ \bibinfo {author} {\bibfnamefont {Yaroslav~M}\ \bibnamefont
			{Blanter}},\ }\href@noop {} {\emph {\bibinfo {title} {Quantum Transport:
				Introduction to Nanoscience}}}\ (\bibinfo  {publisher} {Cambridge University
		Press},\ \bibinfo {year} {2009})\BibitemShut {NoStop}%
	\bibitem [{\citenamefont {Krause}\ \emph
		{et~al.}(2022{\natexlab{a}})\citenamefont {Krause}, \citenamefont {Brovang},
		\citenamefont {Rasmussen}, \citenamefont {Chatterjee},\ and\ \citenamefont
		{Kuemmeth}}]{Krause2021}%
	\BibitemOpen
	\bibfield  {author} {\bibinfo {author} {\bibfnamefont {Oswin}\ \bibnamefont
			{Krause}}, \bibinfo {author} {\bibfnamefont {Bertram}\ \bibnamefont
			{Brovang}}, \bibinfo {author} {\bibfnamefont {Torbjørn}\ \bibnamefont
			{Rasmussen}}, \bibinfo {author} {\bibfnamefont {Anasua}\ \bibnamefont
			{Chatterjee}}, \ and\ \bibinfo {author} {\bibfnamefont {Ferdinand}\
			\bibnamefont {Kuemmeth}},\ }\bibfield  {title} {\enquote {\bibinfo {title}
			{Estimation of convex polytopes for automatic discovery of charge state
				transitions in quantum dot arrays},}\ }\href@noop {} {\bibfield  {journal}
		{\bibinfo  {journal} {Electronics}\ }\textbf {\bibinfo {volume} {11}},\
		\bibinfo {pages} {15} (\bibinfo {year} {2022}{\natexlab{a}})}\BibitemShut
	{NoStop}%
	\bibitem [{\citenamefont {Gaudreau}\ \emph {et~al.}(2011)\citenamefont
		{Gaudreau}, \citenamefont {Granger}, \citenamefont {Kam}, \citenamefont
		{Aers}, \citenamefont {Studenikin}, \citenamefont {Zawadzki}, \citenamefont
		{Pioro-Ladri{\`{e}}re}, \citenamefont {Wasilewski},\ and\ \citenamefont
		{Sachrajda}}]{Gaudreau2011}%
	\BibitemOpen
	\bibfield  {author} {\bibinfo {author} {\bibfnamefont {L.}~\bibnamefont
			{Gaudreau}}, \bibinfo {author} {\bibfnamefont {G.}~\bibnamefont {Granger}},
		\bibinfo {author} {\bibfnamefont {A.}~\bibnamefont {Kam}}, \bibinfo {author}
		{\bibfnamefont {G.~C.}\ \bibnamefont {Aers}}, \bibinfo {author}
		{\bibfnamefont {S.~A.}\ \bibnamefont {Studenikin}}, \bibinfo {author}
		{\bibfnamefont {P.}~\bibnamefont {Zawadzki}}, \bibinfo {author}
		{\bibfnamefont {M.}~\bibnamefont {Pioro-Ladri{\`{e}}re}}, \bibinfo {author}
		{\bibfnamefont {Z.~R.}\ \bibnamefont {Wasilewski}}, \ and\ \bibinfo {author}
		{\bibfnamefont {A.~S.}\ \bibnamefont {Sachrajda}},\ }\bibfield  {title}
	{\enquote {\bibinfo {title} {{Coherent control of three-spin states in a
					triple quantum dot}},}\ }\href@noop {} {\bibfield  {journal} {\bibinfo
			{journal} {Nature Physics}\ }\textbf {\bibinfo {volume} {8}},\ \bibinfo
		{pages} {54--58} (\bibinfo {year} {2011})}\BibitemShut {NoStop}%
	\bibitem [{\citenamefont {Hsiao}\ \emph {et~al.}(2020)\citenamefont {Hsiao},
		\citenamefont {van Diepen}, \citenamefont {Mukhopadhyay}, \citenamefont
		{Reichl}, \citenamefont {Wegscheider},\ and\ \citenamefont
		{Vandersypen}}]{Hsiao2020}%
	\BibitemOpen
	\bibfield  {author} {\bibinfo {author} {\bibfnamefont {T.-K.}\ \bibnamefont
			{Hsiao}}, \bibinfo {author} {\bibfnamefont {C.J.}\ \bibnamefont {van
				Diepen}}, \bibinfo {author} {\bibfnamefont {U.}~\bibnamefont {Mukhopadhyay}},
		\bibinfo {author} {\bibfnamefont {C.}~\bibnamefont {Reichl}}, \bibinfo
		{author} {\bibfnamefont {W.}~\bibnamefont {Wegscheider}}, \ and\ \bibinfo
		{author} {\bibfnamefont {L.M.K.}\ \bibnamefont {Vandersypen}},\ }\bibfield
	{title} {\enquote {\bibinfo {title} {Efficient orthogonal control of tunnel
				couplings in a quantum dot array},}\ }\href@noop {} {\bibfield  {journal}
		{\bibinfo  {journal} {Physical Review Applied}\ }\textbf {\bibinfo {volume}
			{13}},\ \bibinfo {pages} {054018} (\bibinfo {year} {2020})}\BibitemShut
	{NoStop}%
	\bibitem [{\citenamefont {Borjans}\ \emph {et~al.}(2021)\citenamefont
		{Borjans}, \citenamefont {Mi},\ and\ \citenamefont {Petta}}]{Borjans2021}%
	\BibitemOpen
	\bibfield  {author} {\bibinfo {author} {\bibfnamefont {F.}~\bibnamefont
			{Borjans}}, \bibinfo {author} {\bibfnamefont {X.}~\bibnamefont {Mi}}, \ and\
		\bibinfo {author} {\bibfnamefont {J.R.}\ \bibnamefont {Petta}},\ }\bibfield
	{title} {\enquote {\bibinfo {title} {Spin digitizer for high-fidelity readout
				of a cavity-coupled silicon triple quantum dot},}\ }\href {\doibase
		10.1103/PhysRevApplied.15.044052} {\bibfield  {journal} {\bibinfo  {journal}
			{Phys. Rev. Applied}\ }\textbf {\bibinfo {volume} {15}},\ \bibinfo {pages}
		{044052} (\bibinfo {year} {2021})}\BibitemShut {NoStop}%
	\bibitem [{\citenamefont {Jordan}(2010)}]{Jordan2010}%
	\BibitemOpen
	\bibfield  {author} {\bibinfo {author} {\bibfnamefont {Stephen~P}\
			\bibnamefont {Jordan}},\ }\bibfield  {title} {\enquote {\bibinfo {title}
			{{Permutational Quantum Computing}},}\ }\href@noop {} {\bibfield  {journal}
		{\bibinfo  {journal} {Quantum Information and Computation}\ }\textbf
		{\bibinfo {volume} {10}},\ \bibinfo {pages} {470} (\bibinfo {year}
		{2010})}\BibitemShut {NoStop}%
	\bibitem [{\citenamefont {Gaudreau}\ \emph {et~al.}(2006)\citenamefont
		{Gaudreau}, \citenamefont {Studenikin}, \citenamefont {Sachrajda},
		\citenamefont {Zawadzki}, \citenamefont {Kam}, \citenamefont {Lapointe},
		\citenamefont {Korkusi{\'{n}}ski},\ and\ \citenamefont
		{Hawrylak}}]{Gaudreau2006}%
	\BibitemOpen
	\bibfield  {author} {\bibinfo {author} {\bibfnamefont {L.}~\bibnamefont
			{Gaudreau}}, \bibinfo {author} {\bibfnamefont {S.~A.}\ \bibnamefont
			{Studenikin}}, \bibinfo {author} {\bibfnamefont {A.~S.}\ \bibnamefont
			{Sachrajda}}, \bibinfo {author} {\bibfnamefont {P.}~\bibnamefont {Zawadzki}},
		\bibinfo {author} {\bibfnamefont {A.}~\bibnamefont {Kam}}, \bibinfo {author}
		{\bibfnamefont {J.}~\bibnamefont {Lapointe}}, \bibinfo {author}
		{\bibfnamefont {M.}~\bibnamefont {Korkusi{\'{n}}ski}}, \ and\ \bibinfo
		{author} {\bibfnamefont {P.}~\bibnamefont {Hawrylak}},\ }\bibfield  {title}
	{\enquote {\bibinfo {title} {Stability diagram of a few-electron triple
				dot},}\ }\href@noop {} {\bibfield  {journal} {\bibinfo  {journal} {Physical
				Review Letters}\ }\textbf {\bibinfo {volume} {97}},\ \bibinfo {pages}
		{036807} (\bibinfo {year} {2006})}\BibitemShut {NoStop}%
	\bibitem [{\citenamefont {Hamo}\ \emph {et~al.}(2016)\citenamefont {Hamo},
		\citenamefont {Benyamini}, \citenamefont {Shapir}, \citenamefont {Khivrich},
		\citenamefont {Waissman}, \citenamefont {Kaasbjerg}, \citenamefont {Oreg},
		\citenamefont {von Oppen},\ and\ \citenamefont {Ilani}}]{Hamo2016}%
	\BibitemOpen
	\bibfield  {author} {\bibinfo {author} {\bibfnamefont {A.}~\bibnamefont
			{Hamo}}, \bibinfo {author} {\bibfnamefont {A.}~\bibnamefont {Benyamini}},
		\bibinfo {author} {\bibfnamefont {I.}~\bibnamefont {Shapir}}, \bibinfo
		{author} {\bibfnamefont {I.}~\bibnamefont {Khivrich}}, \bibinfo {author}
		{\bibfnamefont {J.}~\bibnamefont {Waissman}}, \bibinfo {author}
		{\bibfnamefont {K.}~\bibnamefont {Kaasbjerg}}, \bibinfo {author}
		{\bibfnamefont {Y.}~\bibnamefont {Oreg}}, \bibinfo {author} {\bibfnamefont
			{F.}~\bibnamefont {von Oppen}}, \ and\ \bibinfo {author} {\bibfnamefont
			{S.}~\bibnamefont {Ilani}},\ }\bibfield  {title} {\enquote {\bibinfo {title}
			{Electron attraction mediated by {Coulomb} repulsion},}\ }\href@noop {}
	{\bibfield  {journal} {\bibinfo  {journal} {Nature}\ }\textbf {\bibinfo
			{volume} {535}},\ \bibinfo {pages} {395--400} (\bibinfo {year}
		{2016})}\BibitemShut {NoStop}%
	\bibitem [{\citenamefont {Lent}\ \emph {et~al.}(1993)\citenamefont {Lent},
		\citenamefont {Tougaw}, \citenamefont {Porod},\ and\ \citenamefont
		{Bernstein}}]{Lent1993}%
	\BibitemOpen
	\bibfield  {author} {\bibinfo {author} {\bibfnamefont {C~S}\ \bibnamefont
			{Lent}}, \bibinfo {author} {\bibfnamefont {P~D}\ \bibnamefont {Tougaw}},
		\bibinfo {author} {\bibfnamefont {W}~\bibnamefont {Porod}}, \ and\ \bibinfo
		{author} {\bibfnamefont {G~H}\ \bibnamefont {Bernstein}},\ }\bibfield
	{title} {\enquote {\bibinfo {title} {Quantum cellular automata},}\
	}\href@noop {} {\bibfield  {journal} {\bibinfo  {journal} {Nanotechnology}\
		}\textbf {\bibinfo {volume} {4}},\ \bibinfo {pages} {49} (\bibinfo {year}
		{1993})}\BibitemShut {NoStop}%
	\bibitem [{\citenamefont {Amlani}\ \emph {et~al.}(1999)\citenamefont {Amlani},
		\citenamefont {Orlov}, \citenamefont {Toth}, \citenamefont {Bernstein},
		\citenamefont {Lent},\ and\ \citenamefont {Snider}}]{Amlani1999}%
	\BibitemOpen
	\bibfield  {author} {\bibinfo {author} {\bibfnamefont {Islamshah}\
			\bibnamefont {Amlani}}, \bibinfo {author} {\bibfnamefont {Alexei~O.}\
			\bibnamefont {Orlov}}, \bibinfo {author} {\bibfnamefont {Geza}\ \bibnamefont
			{Toth}}, \bibinfo {author} {\bibfnamefont {Gary~H.}\ \bibnamefont
			{Bernstein}}, \bibinfo {author} {\bibfnamefont {Craig~S.}\ \bibnamefont
			{Lent}}, \ and\ \bibinfo {author} {\bibfnamefont {Gregory~L.}\ \bibnamefont
			{Snider}},\ }\bibfield  {title} {\enquote {\bibinfo {title} {Digital logic
				gate using quantum-dot cellular automata},}\ }\href@noop {} {\bibfield
		{journal} {\bibinfo  {journal} {Science}\ }\textbf {\bibinfo {volume}
			{284}},\ \bibinfo {pages} {289--291} (\bibinfo {year} {1999})}\BibitemShut
	{NoStop}%
	\bibitem [{\citenamefont {T\'oth}\ and\ \citenamefont {Lent}(2001)}]{toth2001}%
	\BibitemOpen
	\bibfield  {author} {\bibinfo {author} {\bibfnamefont {G\'eza}\ \bibnamefont
			{T\'oth}}\ and\ \bibinfo {author} {\bibfnamefont {Craig~S.}\ \bibnamefont
			{Lent}},\ }\bibfield  {title} {\enquote {\bibinfo {title} {Quantum computing
				with quantum-dot cellular automata},}\ }\href@noop {} {\bibfield  {journal}
		{\bibinfo  {journal} {Phys. Rev. A}\ }\textbf {\bibinfo {volume} {63}},\
		\bibinfo {pages} {052315} (\bibinfo {year} {2001})}\BibitemShut {NoStop}%
	\bibitem [{\citenamefont {Krause}\ \emph
		{et~al.}(2022{\natexlab{b}})\citenamefont {Krause}, \citenamefont
		{Chatterjee}, \citenamefont {Kuemmeth},\ and\ \citenamefont {van
			Nieuwenburg}}]{Krause2022}%
	\BibitemOpen
	\bibfield  {author} {\bibinfo {author} {\bibfnamefont {Oswin}\ \bibnamefont
			{Krause}}, \bibinfo {author} {\bibfnamefont {Anasua}\ \bibnamefont
			{Chatterjee}}, \bibinfo {author} {\bibfnamefont {Ferdinand}\ \bibnamefont
			{Kuemmeth}}, \ and\ \bibinfo {author} {\bibfnamefont {Evert}\ \bibnamefont
			{van Nieuwenburg}},\ }\bibfield  {title} {\enquote {\bibinfo {title}
			{{Learning Coulomb Diamonds in Large Quantum Dot Arrays}},}\ }\href@noop {}
	{\bibfield  {journal} {\bibinfo  {journal} {SciPost Phys.}\ }\textbf
		{\bibinfo {volume} {13}},\ \bibinfo {pages} {084} (\bibinfo {year}
		{2022}{\natexlab{b}})}\BibitemShut {NoStop}%
	\bibitem [{\citenamefont {Boter}\ \emph {et~al.}(2022)\citenamefont {Boter},
		\citenamefont {Dehollain}, \citenamefont {van Dijk}, \citenamefont {Xu},
		\citenamefont {Hensgens}, \citenamefont {Versluis}, \citenamefont {Naus},
		\citenamefont {Clarke}, \citenamefont {Veldhorst}, \citenamefont
		{Sebastiano},\ and\ \citenamefont {Vandersypen}}]{Boter2021}%
	\BibitemOpen
	\bibfield  {author} {\bibinfo {author} {\bibfnamefont {Jelmer~M.}\
			\bibnamefont {Boter}}, \bibinfo {author} {\bibfnamefont {Juan~P.}\
			\bibnamefont {Dehollain}}, \bibinfo {author} {\bibfnamefont {Jeroen~P.G.}\
			\bibnamefont {van Dijk}}, \bibinfo {author} {\bibfnamefont {Yuanxing}\
			\bibnamefont {Xu}}, \bibinfo {author} {\bibfnamefont {Toivo}\ \bibnamefont
			{Hensgens}}, \bibinfo {author} {\bibfnamefont {Richard}\ \bibnamefont
			{Versluis}}, \bibinfo {author} {\bibfnamefont {Henricus~W.L.}\ \bibnamefont
			{Naus}}, \bibinfo {author} {\bibfnamefont {James~S.}\ \bibnamefont {Clarke}},
		\bibinfo {author} {\bibfnamefont {Menno}\ \bibnamefont {Veldhorst}}, \bibinfo
		{author} {\bibfnamefont {Fabio}\ \bibnamefont {Sebastiano}}, \ and\ \bibinfo
		{author} {\bibfnamefont {Lieven~M.K.}\ \bibnamefont {Vandersypen}},\
	}\bibfield  {title} {\enquote {\bibinfo {title} {Spiderweb array: A sparse
				spin-qubit array},}\ }\href@noop {} {\bibfield  {journal} {\bibinfo
			{journal} {Phys. Rev. Applied}\ }\textbf {\bibinfo {volume} {18}},\ \bibinfo
		{pages} {024053} (\bibinfo {year} {2022})}\BibitemShut {NoStop}%
	\bibitem [{\citenamefont {Terhal}(2015)}]{Terhal2015}%
	\BibitemOpen
	\bibfield  {author} {\bibinfo {author} {\bibfnamefont {Barbara~M.}\
			\bibnamefont {Terhal}},\ }\bibfield  {title} {\enquote {\bibinfo {title}
			{Quantum error correction for quantum memories},}\ }\href {\doibase
		10.1103/RevModPhys.87.307} {\bibfield  {journal} {\bibinfo  {journal} {Rev.
				Mod. Phys.}\ }\textbf {\bibinfo {volume} {87}},\ \bibinfo {pages} {307--346}
		(\bibinfo {year} {2015})}\BibitemShut {NoStop}%
	\bibitem [{\citenamefont {Henk}\ \emph {et~al.}(2017)\citenamefont {Henk},
		\citenamefont {Richter-Gebert},\ and\ \citenamefont {Ziegler}}]{Henk2017}%
	\BibitemOpen
	\bibfield  {author} {\bibinfo {author} {\bibfnamefont {M}~\bibnamefont
			{Henk}}, \bibinfo {author} {\bibfnamefont {J}~\bibnamefont {Richter-Gebert}},
		\ and\ \bibinfo {author} {\bibfnamefont {G~M}\ \bibnamefont {Ziegler}},\
	}\bibfield  {title} {\enquote {\bibinfo {title} {Basic properties of convex
				polytopes},}\ }\href@noop {} {\bibfield  {journal} {\bibinfo  {journal}
			{Handbook of discrete and computational geometry}\ } (\bibinfo {year}
		{2017})}\BibitemShut {NoStop}%
	\bibitem [{\citenamefont {Diamond}\ and\ \citenamefont
		{Boyd}(2016)}]{diamond2016cvxpy}%
	\BibitemOpen
	\bibfield  {author} {\bibinfo {author} {\bibfnamefont {Steven}\ \bibnamefont
			{Diamond}}\ and\ \bibinfo {author} {\bibfnamefont {Stephen}\ \bibnamefont
			{Boyd}},\ }\bibfield  {title} {\enquote {\bibinfo {title} {{CVXPY}: {A}
				{P}ython-embedded modeling language for convex optimization},}\ }\href@noop
	{} {\bibfield  {journal} {\bibinfo  {journal} {Journal of Machine Learning
				Research}\ }\textbf {\bibinfo {volume} {17}},\ \bibinfo {pages} {1--5}
		(\bibinfo {year} {2016})}\BibitemShut {NoStop}%
	\bibitem [{\citenamefont {Agrawal}\ \emph {et~al.}(2018)\citenamefont
		{Agrawal}, \citenamefont {Verschueren}, \citenamefont {Diamond},\ and\
		\citenamefont {Boyd}}]{agrawal2018rewriting}%
	\BibitemOpen
	\bibfield  {author} {\bibinfo {author} {\bibfnamefont {Akshay}\ \bibnamefont
			{Agrawal}}, \bibinfo {author} {\bibfnamefont {Robin}\ \bibnamefont
			{Verschueren}}, \bibinfo {author} {\bibfnamefont {Steven}\ \bibnamefont
			{Diamond}}, \ and\ \bibinfo {author} {\bibfnamefont {Stephen}\ \bibnamefont
			{Boyd}},\ }\bibfield  {title} {\enquote {\bibinfo {title} {A rewriting system
				for convex optimization problems},}\ }\href@noop {} {\bibfield  {journal}
		{\bibinfo  {journal} {Journal of Control and Decision}\ }\textbf {\bibinfo
			{volume} {5}},\ \bibinfo {pages} {42--60} (\bibinfo {year}
		{2018})}\BibitemShut {NoStop}%
	\bibitem [{\citenamefont {Domahidi}\ \emph {et~al.}(2013)\citenamefont
		{Domahidi}, \citenamefont {Chu},\ and\ \citenamefont
		{Boyd}}]{Domahidi2013ecos}%
	\BibitemOpen
	\bibfield  {author} {\bibinfo {author} {\bibfnamefont {A.}~\bibnamefont
			{Domahidi}}, \bibinfo {author} {\bibfnamefont {E.}~\bibnamefont {Chu}}, \
		and\ \bibinfo {author} {\bibfnamefont {S.}~\bibnamefont {Boyd}},\ }\bibfield
	{title} {\enquote {\bibinfo {title} {{ECOS}: {A}n {SOCP} solver for embedded
				systems},}\ }in\ \href@noop {} {\emph {\bibinfo {booktitle} {European Control
				Conference (ECC)}}}\ (\bibinfo {year} {2013})\ pp.\ \bibinfo {pages}
	{3071--3076}\BibitemShut {NoStop}%
	\bibitem [{\citenamefont {Barber}\ \emph {et~al.}(1996)\citenamefont {Barber},
		\citenamefont {Dobkin},\ and\ \citenamefont
		{Huhdanpaa}}]{barber1996quickhull}%
	\BibitemOpen
	\bibfield  {author} {\bibinfo {author} {\bibfnamefont {C~Bradford}\
			\bibnamefont {Barber}}, \bibinfo {author} {\bibfnamefont {David~P}\
			\bibnamefont {Dobkin}}, \ and\ \bibinfo {author} {\bibfnamefont {Hannu}\
			\bibnamefont {Huhdanpaa}},\ }\bibfield  {title} {\enquote {\bibinfo {title}
			{The quickhull algorithm for convex hulls},}\ }\href@noop {} {\bibfield
		{journal} {\bibinfo  {journal} {ACM Transactions on Mathematical Software
				(TOMS)}\ }\textbf {\bibinfo {volume} {22}},\ \bibinfo {pages} {469--483}
		(\bibinfo {year} {1996})}\BibitemShut {NoStop}%
\end{thebibliography}
\end{document}